\documentclass[oneside,11pt]{article}

\usepackage{tgpagella}
\usepackage[english]{babel}    
\usepackage{enumerate} 

\usepackage{multirow}
\usepackage{graphics,latexsym,amsfonts}       
\usepackage{amssymb,amsthm,hyperref,mathtools}   
\usepackage[dvipsnames]{xcolor} 
\usepackage{mathrsfs} 
\usepackage{graphicx} 
\usepackage{float} 
\usepackage{ulem}    
\usepackage{picture}
\hypersetup{
    colorlinks, 
    linkcolor={red!50!black},  
    citecolor={blue!50!black},  
    urlcolor={blue!80!black}   
}
\usepackage[longnamesfirst]{natbib}    
\usepackage{tabularx}

\def\reference#1{\href{#1}{Cliquer ici pour voir une r\'ef\'erence.}} 
\usepackage{sectsty} 

\usepackage{setspace}
\usepackage{caption}
\sectionfont{\normalfont\scshape\centering}
\subsectionfont{\centering}
\providecommand{\U}[1]{\protect \rule{.1in}{.1in}}
{\normalfont\itshape}

\usepackage[top=1.25in,bottom=1.25in,left=1.25in]{geometry}
\def\reference#1{\href{#1}{Click to see a reference}} 

\def\vs{\vspace{-0,1cm}}

\def\X{\mathscr{X}}
\def\es{\varnothing}

\makeatletter

\newtheoremstyle{mytheoremstyle} 
    {\topsep}                    
    {\topsep}                    
    {\itshape}                   
    {}                           
    {\scshape}                   
    {.}                          
    {.5em}                       
    {}  

\theoremstyle{mytheoremstyle}
\newtheorem{theorem}{Theorem} 
\newtheorem*{theorem*}{Theorem}

\newtheorem{lemma}{Lemma}
\newtheorem{corollary}{Corollary} 
\newtheorem*{corollary*}{Corollary} 
\newtheoremstyle{mydefinitionstyle} 
    {\topsep}                    
    {\topsep}                    
    {}                   
    {}                           
    {\scshape}                   
    {.}                          
    {.5em}                       
    {}  
\theoremstyle{mydefinitionstyle}
\newtheorem{definition}{Definition} 
\newtheorem{example}{Example} 

\newtheorem*{question*}{Question}

\newtheorem{remark}{Remark}

\makeatletter

\title{\bf { A measure of choice irrationality based on opposite judgments}\thanks{The author wishes to thank Gennaro Anastasio, Davide Carpentiere, Jean-Paul Doignon, Paolo Ghirardato, Alfio Giarlotta, M.\,Ali Khan, Paola Manzini, Marco Mariotti, Daniele Pennesi, Ernesto Savaglio, Lorenzo Stanca, and Ester Sudano for several comments and suggestions.
The author declares that he has no known competing financial interests or personal relationships that could have appeared to influence the work reported in this paper.
The author received no financial support for the research, authorship, and/or publication of this article.}}

\author{ 
\textsc{Angelo Enrico Petralia}\thanks{University of Catania, Catania, Italy. angelo.petralia@unict.it}  
}

\usepackage{fancyhdr} 
\date{}

\begin{document}
\sloppy 
\maketitle
\begin{abstract}
In many choice settings the decision maker (DM) adopts a criterion which is a mediation between her preference and its opposite.
According to such compromise, the first $i$ alternatives on top of the DM's taste are evaluated according to its negation.
This pattern allows to define the \textit{compromise-based degree of irrationality}, which quantifies the extent of the mediation embraced by the DM in her choice.
Necessary and sufficient conditions characterizing this index are singled out.
I investigate non rationalizable choices that display the lowest degree of irrationality, and I fully identify the preferences that explain the DM's picks by a minimal mediation between opposite judgments.
These datasets account for some well known selection biases, such as second-best procedures, and the handicapped avoidance.
I offer a simple characterization of the choice behavior that exhibits the most severe compromise, and I show that this subclass comprises almost all choices.
Finally, some alternative measures of compromise are proposed, and compared with the score previously determined.

\medskip


\medskip
	
\end{abstract}

\section*{Introduction}

{  Individual choices are typically affected by a tension between conflicting judgments.
Indeed, often the decision maker (DM) is torn between a preference and the opposite criterion.
For instance, as reported by many works on temptation and self-control \citep{GulPesendorfer2004,Noor2011,NoorTakeoka2015}, food habits are determined by a mediation between the DM's preference for tasty meals (the tempting ones) and her desire to diet.
Especially when temptation is bad, as in the framework of \cite{RavidSteverson2021}, these two criteria are represented by inverse rankings over the alternatives.
The same conflict arises between selfishness and reciprocity: the maximization of one's own utility goes against that of the others, mainly in the experimental  settings with zero-sum games outcomes.
However, dictator, gift exchange, and public good experiments show that people often do not pursue their own profit, but they tend to partially reduce it to provide some reward for the others \citep{FehrCharness2025}.
In these situations, the DM's choice is determined by a preference conciliating her gains, and her social concerns, and resolving the dissonance between these two contradictory evaluation of the options. 
This paradigm applies also to guilt aversion \citep{BattigalliDufwenberg2007,EllingsenJohannessonTjottaTorsvik2010,BellemareSebaldSuetens2017}, and self-punishment \citep{FrieheHippelSchielke2021}, behavioral phenomena determined by the contrast between personal utility and its negation. 
Opposite judgments offer a natural explanation also of the role of self-esteem and confidence in choices, investigated by, among others, \cite{ChuangChengChangChiang2013}, and \cite{KoszegiLoewensteinMurooka2022}.
These authors show that people with a low self-view might refuse to achieve the desired goals or risky options, and prefer alternatives which provide little satisfaction, but also require less effort and ability.
As I shall illustrate below, people in these circumstances decide following an offset between the assessment of alternatives determined by full confidence, and a ranking driven by low self-esteem.

In the works mentioned above, the rationalization of the DM's choice is determined by complex utility functions, which depend on several variables describing the observed phenomenon.
However, as remarked in \cite{HarstadSelten2013}, economic agents usually struggle in quantitative decision making, and {  they} may not be efficient maximizers.
{  Moreover, these models are tailored to the specific behavior they cover, and the theoretical settings on which they are grounded  hardly allow for measuring the rationality of the observed choices.
Instead, this approach is general, and it can be applied to different setups. 
}
Indeed, I introduce a multi-self model of choice in which the DM mediates between her preference and the opposite one, by moving the first $i$ alternatives to the bottom of her current evaluation, in reverse order.
This process generates a collection of linear orders, called \textit{compromises on the preference}, which reproduce the amplitude of the DM's trade-off between her judgment and its opposite, and explain the observed choice.
The following examples illustrate such pattern.

\begin{example}[Bad temptation, \citealt{RavidSteverson2021}]\label{EXMP:food_example}
	A DM on diet chooses an afternoon snack. 
	Her refrigerator contains some combination of the following foods: a healthy and unappetizing yogurt (${y}$), a quite unhealthy and inviting sorbet ($s$), and a highly unhealthy and highly tempting ice cream $i$.
	 The DM's choice $c\colon \X\to X$, with  {  $X=\{i,s,y\}$} is defined by 
	 
	 $$i \underline{s} y,\;\; i\underline{s},\;\;i\underline{y},\;\;s\underline{y}.$$
	 
	When the yogurt is paired with the ice-cream or the sorbet, and the sorbet is paired with the ice-cream, the DM follows her diet $\rhd$, defined by $y\rhd s\rhd i.$
	However, when all the alternatives are offered, she is tempted, and, to satisfy a bit her appetite, {  she} follows the compromise $\rhd_1$ such that  $s\rhd_1 i\rhd_1 y$, in which $y$ is disregarded, and $s$ is on top.
			\end{example}

\begin{example}[Self-confidence in risky choices, inspired by the experiments of \citealt{ChuangChengChangChiang2013}]\label{EXMP:projects}
Consider a set $X=\{h,mh,ml,l\}$ of variants of some technological good (a computer, or a car) having respectively high ($h$), medium-high ($mh$), medium-low ($ml$), and low (l) quality and consumer satisfaction.
Such technologies are sophisticated, and demand increasing skills and expertise, but they also generate proportional losses (due to their original price and their maintenance costs), if they break down.
The DM possesses the competences that would allow her to manage any good.
Thus, she aims to obtain the highest satisfaction, as shown by her preference $\rhd$ such that $h\rhd mh\rhd ml \rhd l.$ 
However, if in some situations she feels less confident, she might think she is unable to use $h,$ and maximize the distortion $\rhd_1,$ determined by $mh\rhd_1 ml\rhd_1 l\rhd_1 h,$ to exclude any loss.
Moreover, if her self-confidence is even lower, she may avoid considering $h$ and $mh,$ relegating them to the bottom of her distorted preference $\rhd_2$, defined by $ml\rhd_2 l\rhd_2 mh \rhd_2 h,$ in which $mh$ is better than $h,$ since it brings fewer losses.  
Assume that $c\colon \X\to X$ is defined by 
$$\underline{h}\;mh\;ml\;l, \,\; h\;mh\;\underline{ml}, \,\; h\;\underline{mh}\;l, \,\;  \underline{h}\;ml\;l, \,\; mh\;\underline{ml}\;l, \,\; \underline{h}\;mh, \;\, h\;\underline{ml}, \,\; \underline{h}\;l, \;\, \underline{mh}\;ml, \,\; mh\;\underline{l}, \,\; \underline{ml}\;l.$$

The selections from $h\;mh\;ml\;l,$ $h\;ml\;l,$ $h\;mh,$ $h\;l,$
 and  $ml\;l$ are coherent with the DM's true preference $\rhd$.
Moreover, the DM's choice  from $h\;mh\;l$, $h\;ml,$ and $mh\;ml$ is explained by $\rhd_1.$
Finally, her picks from $h\;mh\;ml,$ $ mh\;ml\;l,$  and $mh\;l$ are obtained by the maximization of $\rhd_2.$
\footnote{The authors propose choice experiments on binary and three-elements sets to establish a positive correlation between self-confidence and the tendency of the subjects to select an \textit{intermediate} option, just as with the menus $h\;mh\;l$ and  $mh\;ml\;l$.}
 \end{example}

This paradigm is not testable, but it naturally yields a \textit{compromise-based degree of irrationality} of a choice, counting the minimum number of alternatives disregarded in the DM's mediation between opposite preferences.    
For instance, in Example \ref{EXMP:food_example} it is evident that $c$ can be justified only adopting a compromise between the DM's diet and her temptation in which at least the most healthy option is downgraded.
Instead, in Example \ref{EXMP:projects} a mediation in which the top-quality option and the second best are neglected is needed.
{  This score takes the DM's preference and the opposite ranking as reference points, enabling the measurement of the rationality of all those decisions influenced by the conflict between two antithetical judgments.} 
Necessary and sufficient observable conditions that characterize this index are singled out, and the linear orders whose compromises explain choice data are partially elicited.   
I investigate non rationalizable choices that exhibit the lowest degree of irrationality, and I identify the preferences that justify them by resorting to a minimum level of compromise.
This subclass explains some well known selection biases, such as second-best procedures, the decoy effect, and the handicapped avoidance.
Finally, I provide a simple characterization of the choice behavior that can be rationalized only by the most severe mediation between the DM's preference and its negation, and I show that it encompasses almost all choices.

I contribute to two strands of the economic literature.
First, the score proposed in this work stems from the assumption that the DM, when she faces some menu, may maximize either her true preference or some other ranking that partially deals with the inverse criterion.
Thus, my analysis falls within the investigation of \textit{multi-self} models of choice \citep{KalaiRubinsteinSpiegler2002,AmbrusRozen2015,GiarlottaPetraliaWatson2022b}, in which the DM's picks are retrieved by the maximization of many preferences.
These flexible methods allow to define \textit{indices of rationality}, which assess the complexity of a dataset usually by counting the number of linear orders needed to justify it.
In Section~\ref{SECT:relation_literature} I compare the existing multi-self models and the derived scores to the compromise-based degree of irrationality.

Second, and more generally, I offer some advancements in measuring the \textit{rationality/irrationality} of observed choice.
Indeed, starting from \cite{Afriat1974}, economists devised techniques to assess the deviations of data from rational standards, either in consumer theory \citep{Varian1990,EcheniqueLeeShum2011,DeanMartin2016,AguiarSerrano2017} 
 or in more abstract settings \citep{ApesteguiaBallester2017a,CarpentiereGiarlottaWatson2023, Caradonna2024,Ribeiro2024}.
I add to this discussion by proposing a gauge of choice irrationality determined by the tension between conflicting judgments, and offering a detailed comparison with some of the existing tests.  
    
The paper is organized as follows. Section~\ref{SECT:preliminaries} collects some preliminary {  concepts}.
In Section~\ref{SECTION:model} I propose a notion of compromise between the DM's preference and its negation.
A collection of linear orders generated by this process explains any observed choice.
In Section~\ref{SECTION:compromise_based_degree_irrationality} I introduce the compromise-based degree of choice irrationality, and I examine its features.
Specifically, Subsection~\ref{SECTION:compromise_based_degree_irrationality}\ref{SUBSECT:General_characterization} contains a general characterization of the score.
In Subsection~\ref{SECTION:compromise_based_degree_irrationality}\ref{SUBSECTION:weakly_harmful_choices} I analyze datasets with \textit{minimal compromise}: these observations are justified by the lowest extent of mediation between the DM's taste and the opposite criterion.
In Subsection~\ref{SECTION:compromise_based_degree_irrationality}\ref{SUBSECTION:irrational_choices_with_maximal_compromise}
I discuss choices  with \textit{maximal compromise}, those revealing the highest degree of irrationality.
In Section~\ref{SECT:relation_literature} I compare my approach with the 
main multi-self rationality measures discussed in the economic literature (Subsection \ref{SECT:relation_literature}\ref{SUBSECT:Multi-self_measures_rationality}), and other relevant indices of choice irrationality (Subsection \ref{SECT:relation_literature}\ref{SUBSECT:Other_rationality_measures}).
{  Section \ref{SECT:alternatives_models_of_compromise} offers an extensive analysis of alternative models of mediation between conflicting judgments, such as \textit{menu-invariant compromise} (Subsection \ref{SECT:alternatives_models_of_compromise}\ref{SUBSECT:menu_invariant_compromise}), \textit{top-2 rationality} (Subsection \ref{SECT:alternatives_models_of_compromise}\ref{SUBSECT:top-2_rationality}), and \textit{nudging} (Subsection \ref{SECT:alternatives_models_of_compromise}\ref{SUBSECT:nudging}).   
Section~\ref{SECT:concluding_remarks} contains some concluding remarks.} 
All the proofs have been collected in the Appendix.}

\section{Preliminaries}\label{SECT:preliminaries}
In what follows, $X$ denotes the \textsl{ground set}, a finite nonempty set of \textit{alternatives}, \textit{items}, or \textit{options}.
A binary relation $\succ$ on $X$ is \textsl{asymmetric} if $x \succ y$ implies $\neg(y \succ x)$, \textsl{transitive} if $x \succ y \succ z$ implies $x \succ z$, and \textsl{complete} if $x \neq y$ implies $x \succ y$ or $y \succ x$ (here $x,y,z$ are arbitrary elements of $X$). 
A \textsl{(strict) linear order} $\rhd$ is an asymmetric, transitive, and complete binary relation.
I denote by $\mathsf{LO}(X)$ the family of all linear orders on $X$.

 Any nonempty set $A \subseteq X$ is a \textsl{menu}, and $\X = 2^X \setminus \{\es\}$ denotes the family of all menus.
A \textsl{choice function} on $X$ is a map $c \colon \mathscr{X}\rightarrow X$ such that $c(A)\in A$ for any $A\in\X$. 
I refer to a choice function as a \textsl{choice}.
To simplify notation, I often omit set delimiters and commas: thus, $A \cup x$ stands for $A \cup \{x\}$, $A\setminus x$ stands for $A\setminus \{x\}$, $c(xy)=x$ for $c(\{x,y\})=x$, etc.

 Given an asymmetric relation $\succ$ on $X$ and a menu $A \in \X$, the set of \textsl{maximal} elements of $A$ is $\max(A,\succ)=\{x \in X : y \succ x \text{ for no } y \in A\}$. 
A choice $c \colon \mathscr{X} \to X$ is \textsl{rationalizable} if there is a linear order $\rhd$ such that, for any $A \in \mathscr{X}$, $c(A)$ is the unique element of the set $\max(A,\rhd)$; 
in this case I write $c(A) = \max(A,\rhd)$.
The rationalizability of a choice is characterized by the \textsl{Weak Axiom of Revealed Preference} \citep{Samuelson1938}, which says that if an alternative $x$ is chosen when $y$ is available, then $y$ cannot be chosen when $x$ is available:

\begin{definition}[\citealt{Samuelson1938}]
	A choice $c\colon\X\to X$ \textit{satisfies the Weak Axiom of Revealed Preference (WARP)} if for all $A,B \in \X$ and $x,y \in X$, if $x,y \in A \cap B$ and $c(A) = x$, then $c(B) \neq y$.
	Alternatively, I say that \textit{WARP holds for c.} 
\end{definition}

Violations of WARP describe features of irrationality.
In this work, I  call them \textit{reversals}.  

\begin{definition} \label{DEF:minimal_violations_of_alpha}
	{  For any choice $c \colon \X \to X$, a \textsl{reversal} is a pair $(A,B)$ of menus such that  $c(A),c(B)\in A\cap B$, and $c(A)\neq c(B)$.}
\end{definition} 

{  Note that for choices WARP is equivalent to the property of \textit{Contraction Consistency} due to \cite{Chernoff1954}, also called \textit{Independence of  Irrelevant Alternatives} (\textit{IIA}) by  \cite{RadnerMarschak1954} and \cite{Ray1973}. 
This axiom states that if an option is chosen, then it is always selected if other alternatives are removed from the menu.

\begin{definition}\label{DEF:IIA}
	A choice $c\colon\X\to X$ \textit{satisfies the Independence of Irrelevant Alternatives (IIA)} if for any $A,B\in\X$ such that $A\subseteq B$ and $c(B)\in A$, the equality $c(A)=c(B)$ holds.
\end{definition}

Thus, violations of IIA are defined as follows.

\begin{definition} \label{DEF:minimal_violations_of_alpha}
	Given a choice function $c \colon \X \to X$ on $X$, a \textsl{violation of IIA} is an ordered pair $(A,B)$ of menus such that $A \subseteq B$ and $c(A) \neq c(B) \in A$. 
\end{definition} 

As expected, each reversal triggers a violation of IIA, and vice versa.

\begin{lemma}
	Let  $c\colon\X\to X$ be a choice on $X$. 
There are $A,B\in\X$ such that $(A,B)$ is a reversal if and only if
		 there are $D,E\in\X$ such that $(D,E)$ is a violation of IIA.\footnote{ The proof of this result is straightforward, and it is left to the reader.}			
\end{lemma}}

\section{Compromises on preference}\label{SECTION:model}

{  I first propose a notion of \textsl{compromise} on individual taste, which encodes the DM's tendency to mediate between her true preference and the opposite ranking.
I need some preliminary notation.
Given a set $X$, and some $0\leq i\leq \vert X\vert-1,$ I denote by $X^{\rhd}_i$ the set of the first $i$ items on top of $X$ with respect to $\rhd.$

\begin{definition}\label{DEF:preference_compromise}

Given a set $X$, some $\rhd\in\mathsf{LO}(X)$, and $0\leq i \leq \vert X\vert-1,$ the \textit{i-th compromise on $\rhd$} is the binary relation, denoted by $\rhd_i$, such that
\begin{enumerate}[\rm(i)]
	
	\item for any $a,b\in X\setminus X^{\rhd}_i$, $a\rhd b$ implies $ a\rhd_i b$, and
	\item for any $a\in X^{\rhd}_i$ and $b\in X$, $a\rhd b$ implies $b\rhd_i a.$ 
\end{enumerate} 
A linear order $\rhd^{\prime}\in\mathsf{LO}(X)$ is \textit{a compromise on $\rhd$} if $\rhd^{\prime}\equiv \rhd_i $ for some $i\in\{0,\cdots,\vert X\vert-1\}.$
I denote by $-\rhd$ the compromise $\rhd_{\vert X\vert-1}$ on $\rhd.$
 I denote by $\mathsf{Comp}(\rhd)$ the family $\{\rhd_i\}_{0\leq\, i\,\leq \vert X\vert-1}$ of all the $\vert X\vert$ compromises on $\rhd$.
\end{definition}

The $i$-th compromise $\rhd_i$ on a preference $\rhd$ is obtained by shifting the top $i$ alternatives to the bottom and then rearranging them in reverse order.
Observe that, for any $\rhd\in\mathsf{LO}(X)$ and each $0\leq i\leq \vert X\vert-1$, the binary relation $\rhd_i$ is a linear order, and it is unique.
Moreover, since $\rhd_{0}\equiv\rhd$, I have that $\rhd\in\mathsf{Comp}(\rhd)$.
Finally, in Definition~\ref{DEF:preference_compromise} I require that $i<\vert X\vert$,  and exclude $\rhd_{\vert X\vert},$ since $\rhd_{\vert X\vert}\equiv \rhd_{\vert X\vert-1}\equiv -\rhd.$

A compromise naturally reproduces the mediation between the DM's evaluation of the alternatives, and the opposite ranking $-\rhd$.
According to this process, the DM partially follows her preference $\rhd$, as indicated by condition (i) of Definition~\ref{DEF:preference_compromise}, which requires that the relation between the alternatives not contained in $X^{\rhd}_i$ does not change.
However, she also deals with the antithetical criterion $-\rhd$, since, as indicated by condition (ii) the items belonging to $X^{\rhd}_i$ are downgraded, and { they} hold in the compromise $\rhd_i$ the same position they occupy in $-\rhd.$ 
Thus, the DM conciliates her judgment with the opposite one, by prioritizing, to some extent, the alternatives of $X$ that are decently ranked according to $\rhd,$ but are not excessively penalized by $-\rhd.$
Note that, when $i=0,$ the DM adopts her true preference.
However, as $i$ becomes closer to $\vert X\vert-1$, the adversarial criterion is increasingly favored.
Indeed, condition (ii) implies that the best $i$  options are worse than
any other object, even those that are at the peak of the opposite judgment.
For instance, in Example~\ref{EXMP:food_example}, according to the compromise $\rhd_1$, the yogurt, which is the healthiest product, is preferred to the ice cream, the most tempting alternative.
Similarly, in Example~\ref{EXMP:projects} the compromise $\rhd_2$ ranks the worst object better than the high and medium-high quality ones, since the DM aims to avoid any loss from her purchase.
Finally, when $i$ equals $\vert X\vert-1$, the DM experiences the most severe compromise, and embraces the negation of her preference.
I now consider a choice behavior affected by preference compromise

\begin{definition}\label{DEF:rationalization_by_compromise}
	Let  $c\colon \X \to X$ be a choice function on a set $X$, and $\rhd$ be a linear order in $\mathsf{LO}(X)$.
		A family $\mathsf{Comp}_c(\rhd)\subseteq \mathsf{Comp}(\rhd)$ is a \textit{rationalization by compromise on $\rhd$ of $c$} if for every $A\in\X$ there is $\rhd_i\in\mathsf{Comp}_c(\rhd)$ such that $c(A)=\max(A,\rhd_i)$.
\end{definition}

A rationalization by compromise justifies the observed behavior by the DM's attempts to reconcile conflicting assessments.
When she faces a menu, she decides maximizing some compromise on her preference.
Note that in the model described in Definition~\ref{DEF:rationalization_by_compromise} the DM's mediation is menu-dependent.
{  This assumption ensures that the proposed theoretical setting provides justification for all observable choices, thereby establishing the measure of rationality that will be analyzed in this work.}  
Moreover, the tension between opposite preferences affects the evaluation of all the alternatives (those belonging to the ground set), and not only the order of the available options.
The economic interpretation of such assumption is simple: either the DM has already been informed of all the options in the ground set at the beginning of the experiment, or, in an empirical setting, she has a clear sense of the alternatives that are usually available.
Thus, even if compromise determines her selection from a menu, it acts on the DM's mind, shaping her overall ranking.\footnote{In Section~\ref{SECT:alternatives_models_of_compromise} I also discuss a \textit{menu-invariant} version of the method proposed in Definition~\ref{DEF:rationalization_by_compromise}, and I characterize it.}

\begin{remark}\label{REM:menu_dependent_compromise}
	{  A compromise on the DM's preference can be also interpreted as a \textit{satisficing} behavior, in which the DM favors the alternative providing her some level of utility.
	In light of this, given a choice $c\colon \X\to X$, one can define a   \textit{rationalization} \textit{by satisfaction of} $c$ as a pair $\left(\rhd, I\right)$, with $\rhd\in\mathsf{LO}(X)$ and $I\subseteq \{0,\cdots,\vert X\vert-1\}$, such that for any $A\in\X$ there is $i\in I$ for which $A\setminus A^{\rhd}_i\neq \es$ (i.e. if $\vert A\vert\geq i+1$) and $c(A)=\max(A\setminus A^{\rhd}_i,\rhd)$ hold.
	According to this rule, the DM picks the $(i+1)$-th option of a menu, for some $i\in I$.
	The indices of a rationalization by satisfaction and those { denoting} the compromises rationalizing the same dataset are generally distinct.
	However, as shown in Subsection~\ref{SECT:alternatives_models_of_compromise}\ref{SUBSECT:menu_invariant_compromise}, the menu-invariant versions of these two paradigms, in which the utility threshold and and the extent of mediation are fixed, admit representations of an observed choice based on the same index.} 
\end{remark}    

The approach proposed in Definition~\ref{DEF:rationalization_by_compromise} is a special case of the \textsl{rationalization by multiple rationales} proposed by \cite{KalaiRubinsteinSpiegler2002}.
According to their paradigm, the DM is allowed to use several linear orders to justify her choice: she selects from each menu the element that is maximal according to some of these preferences. %
Definition~\ref{DEF:rationalization_by_compromise} is more binding: a rationalization by compromise explains a choice only by means of linear orders which are all compromises on a given preference.
However, if I do not impose any further restriction on the family of compromises that are needed to justify choice data,  my approach cannot be tested.

\begin{lemma}\label{LEMMA:RSP_non_testable_method}
Let   $c\colon	\X\to X$ be a choice on $X$.
For any linear order $\rhd\in \mathsf{LO}(X)$, there is a rationalization by compromise $\mathsf{Comp}_c(\rhd)$ on $\rhd$ of $c$.	
\end{lemma}
 
Even if the method described in Definition~\ref{DEF:rationalization_by_compromise} has no empirical power, it allows the experimenter to infer the extent of the DM's mediation from choice data, as shown in the next section.}

\section{Compromise-based degree of irrationality}\label{SECTION:compromise_based_degree_irrationality}

{Compromise naturally yields a measure of choice irrationality, obtained by estimating the maximum number of preferred alternatives that the DM neglected.
    
\begin{definition}\label{DEF:compromise_based_degree_irrationality}
	 Given a choice $c\colon\X\to X$, I denote by
	$$irr_{\mathsf{Comp}}(c)=\min_{\rhd\in\mathsf{LO}(X)}\left(\min_{\mathsf{Comp}_c(\rhd)\subseteq \mathsf{Comp(\rhd)}}\left(\max_{i\colon\rhd_i\in\mathsf{Comp}_c(\rhd)} i\right)\right)$$
	the \textsl{compromise-based degree of irrationality} of $c$.
\end{definition}

The score proposed in Definition~\ref{DEF:compromise_based_degree_irrationality} is a lower bound, among all the rationalizations by compromise of the observed choice, to the maximum index of the compromise on the DM's true preference.
It encodes the irrationality of a choice as the extent of her mediation between her judgment and the opposite one,  by counting  maximum number of alternatives the DM must have disregarded in her choice.
Note that, by Definition~\ref{DEF:preference_compromise} and Lemma~\ref{LEMMA:RSP_non_testable_method}, $0\leq irr_{\mathsf{Comp}}(c)\leq \vert X\vert-1$ for any choice $c$ defined on a ground set $X$.
Moreover, $irr_{\mathsf{Comp}}(c)=0$ for any rationalizable choice $c$.
When $irr_{\mathsf{Comp}}(c)$ is low, or equal to $0$, the DM evaluates the available options according to her true preference, and in some occasions she maximizes compromises on her preferences in which she neglects few alternatives.
When $irr_{\mathsf{Comp}}$ is closer or equal to $\vert X\vert-1$, the DM embraces a \textit{severe} mediation between her taste and its negation, adopting in her selection compromises in which many alternatives have been downgraded.

\begin{remark}
	{  The compromise-based degree of irrationality depends on the opposite ranking, and the notion of mediation previously defined.
	Indeed, compromises cover the entire spectrum of trade-offs, as formalized in Definition \ref{DEF:preference_compromise},  between $\rhd$ and $-\rhd$, by moving the first $i$ alternatives on top of $\rhd$ to the same position the hold in $-\rhd$, and $irr_{\mathsf{Comp}}$ is a lower bound to the maximum value of $i$ needed to rationalize the dataset.
Thus, a variant of such score grounded on the duality between the DM's taste and a ranking different from its negation can be determined only by either defining a different distortion of the DM's preference, or considering a less extreme compromise, whose $i$ is lower than $\vert X\vert-1.$
 The former task is partly  accomplished in Section \ref{SECT:alternatives_models_of_compromise}, while the latter would be formally equivalent to setting an upper bound to $irr_{\mathsf{Comp}}$, thereby preventing the assessment of the irrationality of some choices. 
	Indeed, given a choice $c\colon\X\to X$, $\rhd\in\mathsf{LO}(X)$, some $j\in\{0,\cdots,\vert X\vert-2\}$, and denoted by $\mathsf{Comp}^{\,j}_c(\rhd)$ all the rationalizations by compromise of $c$ by $\rhd$ that contain only compromises of index at most equal to $j$, it is straightforward to show that 
	
$$\min_{\rhd\in\mathsf{LO}(X)}\left(\min_{\mathsf{Comp}^{\,j}_c(\rhd)\subseteq \mathsf{Comp(\rhd)}}\left(\max_{i\colon\rhd_i\in\mathsf{Comp}^{\,j}_c(\rhd)} i\right)\right),$$

if it exists, it is equal to $$\min_{\rhd\in\mathsf{LO}(X)}\left(\min_{\mathsf{Comp}_c(\rhd)\subseteq \mathsf{Comp(\rhd)}}\left(\max_{i\colon\rhd_i\in\mathsf{Comp}_c(\rhd)\,\wedge\,i\,\leq\,j} i \right)\right),$$

that is, the value that $irr_{\mathsf{Comp}}(c)$ assumes when $irr_{\mathsf{Comp}}(c)\leq j$.}
\end{remark}

The computation of the compromise-based degree of irrationality for few alternatives is not demanding, as shown in the following example.

\begin{example}\label{EXAMPLE:computation_degree_self_punishment}
	Let $c\colon\X\to X$ be  defined on $X=\{x,y,z\}$ as follows:
	$$\underline{x}yz,\;\; x\underline{y},\;\;\underline{x}z,\;\;y\underline{z}.$$

The choice $c$ violates WARP, thus is not rationalizable, and $irr_{\mathsf{Comp}}(c)>0$.
Moreover, the family $\{\rhd\equiv \rhd_0 ,\rhd_1\}$  in which $\rhd$ is such that $x\,\rhd\,z\,\rhd y$, is a rationalization by compromise on $\rhd$  by $c$.
Thus, I conclude that $irr_{\mathsf{Comp}}(c)=1$. 
\end{example}

As the cardinality of the ground set increases, the assessment of $irr_{\mathsf{Comp}}$ by means of Definition~\ref{DEF:compromise_based_degree_irrationality} becomes a complex task.                                                                                                                                                                                                                                                                                                                                                                                                                                                                                                                                                                                                                                                                                                                                                                                                                                                                                                                                                                                                                                                                                                                                                                                                                                                                                                                                                                                                                                                                                                                                                                                                                                                                                                                                                                                                                                                                                                                                                                                                                                                                                                                                                                                                                                                                                                                                                                                                                                                                                                                                                                                                                                                                                                                                                                                                                                                                                                                                                                                                                                                                                                                                                                                                                                                                                                                                                                                                                                                                                                                                                                                                                                                                                                                                                                                                                                                                                                                                                                                                                                                                                                                                                                                                                                                                                                                                                                                                                                                                                                                                                                                                                                                                                                                                                                                                                                                                                                                                                                                                                                                                                                                                                                                                                                                                                                                                                                                                                                                                                                                                                                                                                                                                                                                                                                                                                                                                                                                                                                                                                                                                                                                                                                                                                                                                                                                                                                                                                                                                                                                                                                                                                                                                                                                                                                                                                                                                                                                                                                                                                                                                                                                                                                                                                                                                                                                                                                                                                                                                                                                                                                                                                                                                                                                                                 
However, the compromise-based degree of irrationality can be inferred from some properties of the dataset.  
Indeed, in the next three subsections  I investigate the choice behaviors determined by this rationality measure, and I show that they are characterized by some axioms of choice data.
Moreover, the unobserved preferences, and the compromises on them explaining the observed choices are identified. 
{ Subsection~\ref{SECTION:compromise_based_degree_irrationality}\ref{SUBSECT:General_characterization}} offers a general characterization of the compromise-based degree of irrationality, and a partial elicitation of the DM's true taste.
In { Subsection~\ref{SECTION:compromise_based_degree_irrationality}\ref{SUBSECTION:weakly_harmful_choices}} non rationalizable outcomes exhibiting a minimal mediation are analyzed, and the linear orders generating the compromises that justify the data are uniquely inferred.
Finally, in { Subsection~\ref{SECTION:compromise_based_degree_irrationality}\ref{SUBSECTION:irrational_choices_with_maximal_compromise}}
I assume that the DM adopts the most severe compromise, selecting in some occasions the least preferred alternative.
I provide an alternative and simpler characterization of the choices belonging to this pattern, which, as the size of the ground set grows, prevails.}

\subsection{Characterization and partial identification}\label{SUBSECT:General_characterization}
The scope of the DM's compromise on her preference can be elicited from the following property of the dataset.

\begin{definition}\label{DEF:violation_independent_selection_i_items}
	A choice $c\colon \X\to X$ {\it violates WARP under constant nonreciprocal selection of $j$ items}, with  $j\in\{1,\cdots,\vert X\vert-1\},$  if 
	
		\begin{enumerate}[\rm(i)]
	\item  for any $D\subset X$ of size $\vert D\vert < j$ there is a reversal $(A,B)$ such that $c(A)\neq x\neq c(B)$ for any $x\in D$, and   
	\item there is a (arbitrarily ordered) set $\{x_1,\cdots,x_j\}\subset X$ such that 
	\begin{enumerate}
	\item for any reversal $(A,B)$ either $c(A)=x_h$ or $c(B)=x_h$ holds for some $h\in\{1,\cdots,j\},$ and 
		\item for any $h\in\{1,\cdots,j\}$ there is a reversal $(A,B)$ such that either $x_h=c(A),y=c(B),$ or $y=c(A),x_h=c(B)$ holds, with $y\in X\setminus \{x_1,\cdots,x_{j}\}$.
	\end{enumerate}

	\end{enumerate}
\end{definition} 

 Condition (i) of Definition~\ref{DEF:violation_independent_selection_i_items} imposes that there is no subset of cardinality lower than $j$ whose options are involved in any observed violation of WARP. 
Moreover, according to condition (ii)(a) of the property mentioned above, there are  alternatives $x_1,\cdots,x_j$ which are involved in any reversal.
Finally, (ii)(b) requires that the experimenter be able to find, for each item $x_h\in \{x_1,\cdots,x_j\}$, a reversal $(A,B)$ in which $x_h$ is selected, and another item $y,$ distinct from $x_1,\cdots, x_{j},$ is chosen.
{  This axiom can be easily detected from data, as shown in the next example.

\begin{example}
Given $X=\{w,x,y,z\}$, let $c\colon\X\to X$ be the choice defined by 

$$\underline{w}xyz,\;\; w\underline{x}y,\;\; \underline{w}xz,\;\;\underline{w}yz,\;\; x\underline{y}z, \;\; w\underline{x},\;\; w\underline{y},\;\; \underline{w}z,\;\;x\underline{y}, \;\;  x\underline{z},\;\;\underline{y}z.$$	
\end{example}

This choice violates WARP under constant nonreciprocal selection of $2$ items.
To see why, note that each alternative is not selected in some reversal.
Indeed, $w$ and $z$ are not picked in the reversal $(wxy,xy)$, $x$ is not selected from  $(wxyz,wy)$, $y$ is not chosen from any menu of $(wxyz,wxy)$.
Thus, condition (i) of Definition \ref{DEF:violation_independent_selection_i_items} is satisfied for $j=2$.
Moreover, the set $wx$ satisfies (ii)(a), since there is no reversal in which $y$ and $z$ are both selected.
Finally also (ii)(b) holds, since from $(wxyz,wy)$ the items $w$ and $y$ are picked, while in the reversal $(wxy,xy)$ options $x$ and $y$ are chosen.
}

As announced, the property presented in Definition~\ref{DEF:violation_independent_selection_i_items}  characterizes the compromise-based degree of irrationality.

\begin{theorem}\label{THM:characterization_compromise_based_degree_irrationality}
	Let $c\colon\X\to X$ be a choice on $X$, and $1\leq j\leq \vert X\vert-1$.
	We have that $irr_{\mathsf{Comp}}(c)=j$ if and only if $c$ violates WARP under constant nonreciprocal selection of $j$ items. 
\end{theorem} 
   
Thus, the experimenter can verify that the degree of irrationality of the dataset, determined by the DM's compromise between opposite evaluations, equals $j$ by checking that the observed choice violates WARP under constant nonreciprocal selection of $j$ items.
For instance, the choice $c$ displayed in Example~\ref{EXMP:projects} violates WARP under constant selection of $2$ items, $h$ and $mh,$ since in any reversal either $h$ or $mh$ are chosen, and there are reversals in which $h$ or $mh$ are selected from one of the two menus, and another option, distinct from both of them, is picked in the remaining menu.
Thus, I can conclude that $irr_{\mathsf{Comp}}(c)=2.$  
Moreover, the proof of Theorem~\ref{THM:characterization_compromise_based_degree_irrationality} also shows that DM's preference determining a lower bound to her compromise can be partially retrieved.
{  Before formally introducing this result I need some notation.
Let  $c\colon\X\to X$ be a choice violating WARP under constant nonreciprocal selection of $j$ items, with $1\leq j\leq \vert X\vert-1$, and 
	let $\rhd^c$ the binary relation defined by 
	\begin{itemize}
		\item  $x_{g}\rhd^c x_h$ for any $g,h\in\{1,\cdots,j\}$ such that $g<h,$
		 \item $y\rhd^c z$ for any $y,z\in X\setminus \{x_1,\cdots,x_j\}$ for which there is a menu $A\in\X$ such that $z\in A$ and $y=c(A),$ and
 \item $x_g\rhd^c y$ for any $g\in\{1,\cdots,j\}$ and $y\in X\setminus \{1,\cdots,j\}.$ 
	\end{itemize}}
	
Indeed I have:

\begin{corollary}\label{COR:multiple_minimal_preferences}
	Let  $c\colon\X\to X$ be a choice violating WARP under constant nonreciprocal selection of $j$ items, with $1\leq j\leq \vert X\vert-1$.
Then $\rhd^c$ is asymmetric and transitive, and, for any $\rhd\in\mathsf{LO}(X)$ extending $\rhd^c,$ the family $\{\rhd_i \colon i\in\{1,\cdots,j\}\}\subseteq \mathsf{Comp}(\rhd)$ is a rationalization by compromise on $\rhd$ of $c$.
\end{corollary}   

Note that the ordering of the set $\{x_1,\cdots,x_j\}$ is arbitrary.
Thus, Corollary~\ref{COR:multiple_minimal_preferences} also implies that if $j$ is the compromise-based degree of irrationality of a choice, there are at least $j!$ distinct linear orders, which, together with the compromises on them, up the $j$-th one, justify by compromise it.
However, as shown in the next subsection, when the extent of the DM's mediation between her preference and its negation is low, the identification of her taste is far more accurate.

 \subsection{Minimal compromise}\label{SUBSECTION:weakly_harmful_choices}

I now analyze choices that exhibit the \textit{weakest} form of irrationality, in which the individual relies on her true preference, but occasionally she adopts the compromise downgrading the best alternative.

\begin{definition}\label{DEF:irrational_choices_with_minimal_compromise}
	A choice $c$ is \textsl{irrational with minimal compromise} if $irr_{\mathsf{Comp}}(c)=1$.
\end{definition}

Thus, irrational choices with minimal compromise are non rationalizable datasets explained by mediation only by assuming that the DM adopts in her decision at least the compromise on her preference in which the alternative on top, which is the worst one according to the adversarial assessment, is disregarded.
This is true, for instance, for the choice presented in Example~\ref{EXMP:food_example}, which must be justified by relying at least on the DM's diet, and the compromise that neglects the healthiest option.
In addition to temptation, irrational behavior with minimal compromise encloses many anomalous choices determined by cognitive biases that have been discussed in theoretical and experimental economics.

\begin{example}[Second best procedures, \citealt{BaigentGaertner1996, KalaiRubinsteinSpiegler2002,Banerjee2023}]
\label{EXMP:second_best_procedures}
The DM selects from each menu the alternative holding the second place in her preference.
Given a set $X=\{x,y,z\},$ and $\rhd\in\mathsf{LO}(X)$ such that $x\,\rhd\,y\rhd\, z$, define $c\colon \X\to X$ as
$$x\underline{y}z,\;\; x\underline{y},\;\;x\underline{z},\;\;y\underline{z}.$$

Note that $c$ does not satisfy WARP, thus $irr_{\mathsf{Comp}}(c)>0.$
Moreover, the pair $(\rhd^{\prime},\rhd^{\prime}_1),$ in which $\rhd^{\prime}$ is such that $z\rhd^{\prime} y\,\rhd^{\prime} x $ is a rationalization by compromise on $\rhd^{\prime}$ of $c.$
I conclude that $irr_{\mathsf{Comp}}(c)=1,$ and that $c$ is irrational with minimal compromise.
Note that $\rhd^{\prime}=-\rhd.$
Indeed, given the selection of $z$ from the binary menus, the linear order that minimizes the extent of the mediation is the one antithetical to the DM's mind.\footnote{Example~\ref{EXMP:second_best_procedures} suggests that any second best procedure on three items is an irrational choice with minimal compromise.
However, this is not true for a second best procedure on a ground set $X$ of size $\vert X\vert \geq 5$, as it can be deduced from \citet[Proposition 3]{KalaiRubinsteinSpiegler2002}.}   
\end{example}

\begin{example}[Decoy effect, \citealt{HuberPaynePuto1982}]
\label{EXMP:decoy_effect}	
 
The alternatives belonging to $X=\{x,y,z\}$ can be evaluated according to two different attributes, respectively described by the rankings $\rhd$, defined by $ y\rhd x\rhd z,$ and $\rhd_1$, which satisfies $x\rhd_1 z\rhd_1 y.$
In all binary menus the DM chooses by favoring the first attribute.
However, when all the goods are available, the presence of a product (the \textsl{decoy good}, represented by $z$), which is worse in every respect than another option ($y$), may lead the consumer to select this dominant alternative.
The dataset describing this phenomenon is that {  presented in Example \ref{EXAMPLE:computation_degree_self_punishment}, that is}

$$\underline{x}yz,\, x\underline{y},\,\underline{x}z,\,y\underline{z}\,.$$	

Clearly, $c$ is not rationalizable, and $(\rhd,\rhd_1)$ is a rationalization by compromise of $c,$ thus $irr_{\mathsf{Comp}}(c)=1$.
Note that the second attribute $\rhd_1$, triggered by the presence of $z$ in the menu $xyz$, is a compromise on the first attribute $\rhd$, in which $x$, the best option, is dominated by the other alternatives.

\end{example}

{ 
\begin{example}[Handicapped avoidance, \citealt{SnyderKleckTrentaMentzer, CherepanovFeddersenSandroni2013}]
\label{EXMP:handicapped_avoidance}

Assume that the ground set $X=\{x,y,z\}$ contains three alternatives, which are  watching a movie alone ($x$), watching the same movie with a person on the wheelchair ($y$), and watching a distinct movie alone ($z$).
When the $x$ and $y$ are the only available options, the DM chooses { $y$}, and watches the movie with the handicapped.
However, when the movies are different, from the binary menu containing $y$ and $z,$ she selects $z,$ favoring the alternative movie.
The observed choice is

$$xy\underline{z},\, x\underline{y}, \, \underline{x}z, \, y\underline{z},$$

since I assume that when all the items are feasible, the DM wants to avoid the handicapped.
The DM prefers the movie offered in option $x$ to that proposed in $z,$ but she does not feel comfortable watching a movie with a person with a disability.
Thus, her true preference is described by $\rhd$, which satisfies $x \rhd z \rhd y,$  and it is maximized when she faces the menus $xz$ and $yz.$
However, she tends to mask this desire, and from the menus $xyz$ and $xy$ she decides according to the compromise $\rhd_1,$ which partially enhances the time with the handicapped.\footnote{In other settings lying may be costly, and force the DM to tell the truth, as reported by \cite{AbelerNosenzoRaymond2019}.}\,\footnote{  The message delivered by this example is more general, since it applies to all the situations in which goods differ on different dimensions (in this case, content and company), and the DM has a different rank for each. Then, the compromise can be seen as an  ``aggregation" of different preferences for different features that are bundled in a single good or experience.} 
\end{example}

Irrational choices with minimal compromise can be detected thanks to the following axiom. 
 
\begin{definition}\label{DEF:violation_alpha_constant_selection}
	A choice $c\colon\X\to X$ \textsl {violates WARP under constant selection} if it does not satisfy WARP, and there is an item $x^*\in X$  such that for any reversal $(A,B)$ I have either $x^*=c(A)$, or $x^*=c(B)$.
\end{definition}

A choice violates WARP under constant selection if it exhibits at least a reversal, and there is an item $x^*$ which is  selected in any observed reversal.
This property of data is a special case of that presented in Definition~\ref{DEF:violation_independent_selection_i_items}, which applies for $j=1.$
Thus, Theorem~\ref{THM:characterization_compromise_based_degree_irrationality}  yields

\begin{corollary}\label{THM:characterization_weakly_harmful_choices}
	A choice $c\colon \X\to X$ is irrational with minimal compromise if and only if it violates WARP under constant selection.
		\end{corollary}
		
Moreover, and differently from the general case, the experimenter can almost uniquely identify the DM's taste, by eliciting from data a revealed preference that explains the observed choice by resorting to the least level of mediation.
{  To highlight this fact, I need some additional notation.
Let $c\colon \X\to X$ be a choice that violates WARP under constant selection.
Moreover, let $\{x^{*}_j\}_{j\in J}$, with $J=\{1\}$ or $J=\{1,2\},$ be the set of items such that for any reversal $(A,B),$ and any $j\in J$, either $x_j^{*}=c(A)$ or $x_j^*=c(B)$ holds.
For any $j\in J$, let $\rhd^{c,x^*_j}$ be the binary relation on $X$ such that, for any $y\in X\setminus x^*_j$, I have  $x^*_j\rhd^{c,x^*_j} y$, and, for any distinct $y,z\in X\setminus x^*_j$, 
		$y\rhd^{c,x^*_j} z$ if there is $A\in\X$  such that $x^*_j\not\in A$, $z\in A$, and $y=c(A).$\footnote{Note that $\rhd^{c,x^*_j}$ is a special case of the binary relation exhibited in Corollary~\ref{COR:multiple_minimal_preferences}, which holds when $j=1.$
However, in this theorem I define it in a simpler way so as to facilitate the comprehension of the result.
} 
The following result holds.}

\begin{theorem}\label{THM:identification_preferences_irrational_choices_with_minimal_compromise}
	Let $c\colon \X\to X$ be a choice that violates WARP under constant selection. 
Given some $j\in J$, a pair $\left(\rhd^{c,x^*_j},\rhd^{c,x^*_j}_1\right)$ is a rationalization by compromise on $\rhd^{c,x^*_j}$ of $c$.
Moreover, for any $\rhd\not\in \{\rhd^{c,x^*_j}\}_{j\in J},$  and any rationalization by compromise on $\rhd$ by $c$,	 namely $\mathsf{Comp}_c(\rhd),$ I have that $\max_{i\colon \rhd_i\in\mathsf{Comp}_c(\rhd)}i>1.$  
\end{theorem}

{  When the property presented in  Definition~\ref{DEF:violation_alpha_constant_selection} holds, either there is \textit{only one} alternative which is selected in any observed switch, yielding $J=\{1\}$, or there exist \textit{exactly} two options that are picked in each violation of WARP, that is, $J=\{2\}$.
When the former case occurs, Theorem~\ref{THM:identification_preferences_irrational_choices_with_minimal_compromise} indicates that there is only a preference, which, together with its first compromise, justifies the dataset.
Instead, if the latter arises, there  exist exactly two distinct rankings that determine a rationalization by compromise of the observed choice relying on a minimal mediation.}

\subsection{Maximal compromise}\label{SUBSECTION:irrational_choices_with_maximal_compromise}

 I now describe the class encompassing the most irrational choices, explained only by the most severe mediation between the DM's preference and its negation.

\begin{definition}\label{DEF:strongly_harmful_choices}
	A choice $c\colon \X\to X$ is \textit{irrational with maximal compromise}  if $irr_{\mathsf{Comp}}(c)=\vert X\vert-1$.
\end{definition}

An irrational choice with maximal compromise can be justified in this framework only by assuming that the DM in some situations adopts the judgment opposite to {  her} own.
A dataset belonging to this behavior displays several failures of WARP, since picks from distinct menus can be determined by the maximization of antithetic criteria.  
Indeed, by virtue of Theorem \ref{THM:characterization_compromise_based_degree_irrationality} I have:

\begin{corollary}\label{COR:characterization_irrational_choices_with_maximal_compromise}
	A choice is irrational with maximal compromise if and only if it violates WARP under constant nonreciprocal selection of $\vert X\vert-1$ items.
\end{corollary}

It turns out that, when $j=\vert X\vert-1,$ the property of choices described in Definition \ref{DEF:violation_independent_selection_i_items} is equivalent { to} the following one.

\begin{definition}\label{DEF:inconstistency}
	A choice  $c\colon \X\to X$ is \textit{inconsistent} if for any distinct $x,y\in X$ there is a reversal $(A,B)$ such that  $x=c(A)$, and $y=c(B)$.
\end{definition}

Inconsistent choices do not convey coherent information about the DM's revealed preference: they exhibit, for any pair of options, at least a violation of WARP in which both are selected.  
 As previously stated, I have:
 
 \begin{lemma}\label{LEM:violation_independent_selection_i_items_equivalent_inconsistency}
 	A choice $c\colon \X\to X$ violates WARP under constant nonreciprocal selection of $\vert X\vert-1$ items if and only if it is inconsistent.  \end{lemma}

An immediate consequence of Corollary~\ref{COR:characterization_irrational_choices_with_maximal_compromise} and Lemma~\ref{LEM:violation_independent_selection_i_items_equivalent_inconsistency} is

\begin{corollary}\label{COR:equivalence_irrational_with_maximal_compromise_inconsistent_choices}
	A choice $c\colon\X\to X$ is irrational with maximal compromise if and only if it is inconsistent.
\end{corollary} 

Thus, to check whether a choice  is irrational with maximal compromise, the experimenter only needs to verify that the dataset is inconsistent.

\begin{example}\label{EXAMPLE:existence_inconsistent_choices}
	Let $c\colon\X\to X$ be defined on $X=\{w,x,y,z\}$ as follows:
	
	$$\underline{w}xyz,\;\;\;w\underline{x}y,\;\;\;wx\underline{z},\;\;\;w\underline{y}z,\;\;\;\underline{x}yz,\;\;\;\underline{w}x,\;\;\;\underline{w}y,\;\;\;\underline{w}z,\;\;\;x\underline{y},\;\;\;\underline{x}z,\;\;\;y\underline{z}.$$
The reader can check that $c$ is inconsistent.
Indeed, for any pair of alternatives in $\{w,x,y,z\}$ there is a reversal $(A,B)$ in which both of them are selected.   
Thus, $c$ is irrational with maximal compromise. 
The collection $\{\rhd,\rhd_1,\rhd_3\}$, where $\rhd$ is the linear order on $X$ such that $w\rhd x\rhd y\rhd z${ ,} is a rationalization by compromise on $\rhd$ of $c$.
When she faces the menus $wxyz$, $xyz$, $wx$, $wy$, and $wz$, the DM relies on her true preference $\rhd$.
In other situations she maximizes some compromise on $\rhd$.
For instance, when the menus $wxy$, $wyz$ are offered, she decides following the compromise $\rhd_1$, which reflects her willingness to neglect $w$. 
Finally, in the menus $wxz$, $xy$, and $yz$, she adopts $\rhd_3\equiv -\rhd$, the adversarial preference.
\end{example}

Note that the verification of the property expressed in Definition~\ref{DEF:violation_independent_selection_i_items} when $j=\vert X\vert-1$ is way more complex.
Indeed, first the {  experimenter} must check condition (i), by looking at any $D\subset X$ of cardinality lower than $\vert X\vert-1$. 
Second, he needs to establish the existence of a set $\{x_1,\cdots,x_{\vert X\vert-1}\}$ satisfying conditions (ii)(a) and (ii)(b). 
Irrational choices with maximal compromise display an erratic behavior that, even if it can be characterized, lacks identification.
Indeed, Lemma~\ref{LEMMA:RSP_non_testable_method} and Definition~\ref{DEF:strongly_harmful_choices} imply that an irrational choice with maximal compromise admits a rationalization by compromise on \textit{any} preference.
Moreover, as the size of the ground set increases, irrational behavior with maximal compromise becomes prevalent.
{  To present this result, I need some notation.}
 I denote $\mathscr{P}^{\,\text{IMC}}$ the property of being irrational with maximal compromise.\footnote{Recall that a  \textsl{property of choices} is a set $\mathscr{P}$ of choices that is closed under isomorphism. Equivalently, it is a formula of second-order logic, which involves quantification over elements and sets, {  it} has a symbol for choice, and {  it} is invariant under choice isomorphisms. Therefore, to say that a property $\mathscr{P}$ holds for $c$ means that $c' \in \mathscr{P}$ for all choices $c'$ isomorphic to $c$.}
 Moreover, I denote by $T(X)$ and $T\left(\mathscr{P}^{\,\text{IMC}},X\right)$ respectively the total number of choices on a ground set $X$, and the total number of choices on $X$ satisfying the property $\mathscr{P}^{\,\text{IMC}}$ .
 I have: 

\begin{theorem}\label{THM:ubiquity_choices_irrationa_with_maximal_compromise}
	As $\vert X\vert$ goes to infinity, $\frac{T\left(\mathscr{P}^{\,\text{IMC}},\,X\right)}{T(X)}$ tends to one. 
\end{theorem}

Theorem~\ref{THM:ubiquity_choices_irrationa_with_maximal_compromise} shows that almost all the observed choices can be explained only by assuming that the DM adopts the most severe mediation between her preference and the opposite one, and she maximizes both of these conflicting judgments.
This result is significant in two respects.
First, it formalizes a particular version of a \textit{folk theorem} establishing that irrational choice behavior becomes prevalent as the number of feasible alternatives grows.
Similar results have been provided by \cite{KalaiRubinsteinSpiegler2002}, \cite{GiarlottaPetraliaWatson2022}, \cite{GiarlottaPetraliaWatson2022b}, and \cite{deClippelRozen2024} for other choice procedures and measures of rationality.
A more detailed comparison between Theorem~\ref{THM:ubiquity_choices_irrationa_with_maximal_compromise} and some of these asymptotic results is provided in Section~\ref{SECT:relation_literature}.
Second, the ubiquity of irrational choices with maximal compromise also provides a theoretical support of the preference reversal and instability suffered by consumers that face large assortments \citep{Chernev2003a,Chernev2003b,Chernev2006}.
As a matter of fact, in consumer choices opposite attributes (e.g. price and quality, risk and reliability, advanced technology and ease of use, etc.) lead to conflicting evaluations of the alternatives.
The increasing availability of options \textit{weakens} the DM's preference, driving her to favor a product feature on some occasions, and the opposite on others.

\section{Relation with the literature}\label{SECT:relation_literature}

In this section I compare my score with other rationality measures that have been proposed to classify choice data.

\subsection{Multi-self measures of rationality}
\label{SUBSECT:Multi-self_measures_rationality}

 The compromise-based degree of irrationality is determined by a multi-self model of choice, in which the DM is endowed with multiple preferences, and she can adopt many of them to finalize her picks.
Indeed, as already mentioned in Section~\ref{SECTION:model}, compromise  is a special case of the paradigm proposed by \cite{KalaiRubinsteinSpiegler2002}, in which the DM's choice is justified by an arbitrary collection of linear orders.
 More formally, the authors call a family $L = \{^{1}\rhd, \ldots, ^{p}\rhd\}$ of linear orders on $X$ a \textsl{rationalization by multiple rationales of a choice $c\colon\X\to X$} if, for all $A\in \X$, the equality $c(A)=\max(A,^{i}\rhd)$ holds for some $^{i}\rhd$ in $L$. 
This method naturally yields a measure of rationality, based on the minimum number of rationales required to justify an observed choice. 
Indeed, denoted by $r(c)$ the least number of linear orders needed in { a} rationalization by multiple rationales of a choice $c$, the authors show that i) $1 \leqslant r(c) \leqslant \vert X \vert-1$ for any choice $c$ on $X$, and ii) as $\vert X \vert$ goes to infinity, the fraction of choices with $r(c) = \vert X \vert -1$ tends to $1$.
Differently from their approach, compromise explains a choice  only by using linear orders belonging to the family $\mathsf{Comp}(\rhd)$ of compromises on some preference $\rhd\in\mathsf{LO}(X)$.
Thus, any rationalization by compromise is a rationalization by multiple rationales, but not any rationalization by multiple rationales is a rationalization by compromise.
Moreover, the compromise-based degree of irrationality $irr_{\mathsf{Comp}}(c)$ does not count the \textit{number} of linear orders needed to explain a choice $c$, but it considers the minimum value, among all the possible rationalizations by compromise of $c$, of the \textit{maximal index} of the DM's compromise.
This score also determines a partition of \textit{testable} choice behaviors.  
Indeed, to assess $irr_{\mathsf{Comp}}$, the experimenter only needs to verify axiom \ref{DEF:violation_independent_selection_i_items} on data. 
Instead, in \cite{KalaiRubinsteinSpiegler2002} $r(c)$ is not characterized.   
{  Finally, the compromise-based degree of irrationality offers a finer classification of choices.
To see why, first note that it is straightforward to prove that $r(c)\leq irr_{\mathsf{Comp}}(c)$ for any observed choice $c$.
Second, consider the choices $c$ and $c^{\,\prime}$ displayed in Example~\ref{EXMP:compromise_nudging_degree}.
Since $c$ is not rationalizable, and $irr_{\mathsf{Comp}}(c)=1$, I conclude that $r(c)=2$.
Instead, the family $\{^{1}\rhd,^{2}\rhd\}$, with $^{1}\rhd$ defined by $x \,\,^{1}\!\rhd w\,^{1}\!\rhd y\,^{1}\!\rhd z$ and $^{2}\rhd$ defined by $y \,\,^{2}\!\rhd w\,\,^{2}\!\rhd x\,^{2}\!\rhd z$, is a rationalization by multiple rationales of $c^{\,\prime}$, and, since $c^{\,\prime}$ is not rationalizable, I conclude that $r(c^{\,\prime})=2$.
According to my score, though, $c^{\,\prime}$ is more irrational than $c.$}

However, there is an analogy between these two rationality measures.
As for the framework of \cite{KalaiRubinsteinSpiegler2002}, Theorem~\ref{THM:ubiquity_choices_irrationa_with_maximal_compromise} shows the fraction of choices exhibiting the highest degree of compromise-based irrationality tends to $1$ when $\vert X\vert$ goes to infinity.

\cite{AmbrusRozen2015} analyze utilities aggregation rules that allow to explain choice data and that satisfy  some standard behavioral properties of social choice theory.
Rationality is gauged by the number of distinct selves that must be aggregated to justify a dataset.  
Unlike their paradigm, in my approach the DM cannot combine distinct preferences, and must apply only one linear order in each menu.
However, as discussed in Section \ref{SECTION:model}, compromise already reproduces an integration between a single hedonic dimension, and its negation.
 Moreover, in their work the authors rely on \textit{cardinal} features of the DM's preference, which require additional information on the subjects' taste.
 My pattern, instead, is merely ordinal, and the experimenter identifies the compromises adopted by the DM only from observed choices.

\cite{GiarlottaPetraliaWatson2022b} propose a specification of the rationalization by multiple rationales, called \textit{context-sensitive multi-rationalization}, in which the selection from each menu is determined by the maximization of a linear order associated to an available alternative.
The authors define, as in \cite{KalaiRubinsteinSpiegler2002}, a (context-sensitive) index of rationality that counts the least number of linear orders needed in a context-sensitive multi-rationalization of a choice, and they show that the most irrational class of choice datasets asymptotically prevails.
It can be proved that compromise is a special case of the procedure of  \cite{GiarlottaPetraliaWatson2022b}.
Note also that in such context-sensitive method there is a behavioral law that determines which linear order the DM will employ in her selection from each menu.
In this respect, compromise is \textit{context-free}, since the underlying process that links menus and maximizing linear orders is not specified.
However, unlike the context-sensitive index of rationality mentioned beforehand,  the compromise-based degree of irrationality yields testable models of choice, which can be retrieved from data.

\subsection{Other rationality measures}
\label{SUBSECT:Other_rationality_measures}

The compromise-based degree of irrationality shares common features with other measures of rationality.
In his seminal work, \cite{Afriat1974} proposes a test based on the maximization of a utility function distorted by a parameter of cost efficiency, which indicates how severely, at given prices, the DM's demand departs from expenditure minimization.
Similarly, my index is determined by a family of distorted criteria, describing the compromises between the DM's preference, and its negation.
However, it applies to a more general setting, in which prices are not observed.

In such domain \cite{ApesteguiaBallester2017a} count the \textit{swaps}, i.e. they sum, across all menus, the number of alternatives that must be switched with the selected one to obtain a choice rationalizable by a linear order on the ground set.
Thus, they define the \textit{swap index} as the number of swaps determined by a linear order minimizing such sum.
Likewise, the compromise-based degree of irrationality considers the minimum amount of options that must be ``swapped" from the top to the bottom of the DM's preference, in reverse order, to justify choice data.
However, such modifications affect the DM's ordering of the ground set, involving also unavailable options.
The motivation behind this assumption is simple.
Indeed, I presume that the DM's has a clear depiction of all the options that may be offered, most likely because the alternatives of the ground set have been already shown to him at the beginning of the experiment.  
Thus, I expect that a mediation between her taste and its negation alters the ranking of all the items in \textit{her mind}, and not only those contained in the menu.
{ 
Moreover, the swap index appears to distinguish less sharply the rationality of choices.
An instance of this is illustrated by the datasets in Example \ref{EXMP:compromise_nudging_degree}: the reader can verify that the swap index of $c$ and $c^{\prime}$ is $4$, whereas, according to compromise, $c^{\prime}$ is more irrational than $c$.\footnote{  The steps of the computation of the swap index of $c$ and $c^{\prime}$ are omitted, but they are available upon request; note that the linear order $\rhd\in\mathsf{LO}(X)$ such that $w\rhd x\rhd y\rhd z$ holds minimizes the sum of the swaps in  $c$ and $c^{\prime}$.}
}

\cite{CarpentiereGiarlottaWatson2023} define a degree of irrationality determined by the distance, inspired to the metric of \cite{Klamler2008} lately corrected by \cite{CarpentiereGiarlottaWatson2024}, of the observed choice from a rationalizable one.
As for their index, my score consists of the minimum deviations from a rational standard, which is the DM's ``true" preference.
Nevertheless, its computation, as it can be deduced from Theorem~\ref{THM:characterization_compromise_based_degree_irrationality}, is simpler, since it relies on the violations of WARP, and not on the comparison across all menus between the observed picks and those retrieved from a rationalizable choice.

\cite{Caradonna2024} provides a choice irrationality measure disclosing the relation between cycles observed in the selections from menus. 
Although the definition and the characterization of my score rely on different properties, they offer natural explanation of cyclical picks, even for choices exhibiting a minimum level of compromise-based irrationality, as it is evident from Examples \ref{EXMP:decoy_effect} and \ref{EXMP:handicapped_avoidance} of Subsection~\ref{SECTION:compromise_based_degree_irrationality}\ref{SUBSECTION:weakly_harmful_choices}.

\cite{Ribeiro2024} compares choice rationality on the basis of two equivalent incomplete rankings of choices, namely the \textit{Violation Criterion} and the \textit{Predictive Criterion}.
Both of them depend on the rationalizability of \textit{partial choices}, which are obtained by looking only at some menus, among those observable.
Given a choice $c\colon \X\to X$, and $\mathscr{M}\subseteq \X$, the \textit{partial choice} $c_{\mathscr{M}}\colon \mathscr{M}\to X$ on $\mathscr{M}$ is defined by $c_{\mathscr{M}}(B)=c(B)$ for any $B\in\mathscr{M}.$
The Violation Criterion requires that $c\colon \X\to X$  is \textit{at least as rational as} $c^{\,\prime}\colon \X\to X$, formally denoted by $c\succsim^V_{\,rat} c^{\,\prime}$, if, for any $\mathscr{M}\subseteq \X$, when $c_{\mathscr{M}}$, is not rationalizable, $c^{\,\prime}_{\mathscr{M}}$ is also not rationalizable.
Moreover, $c$ is \textit{more rational than} $c^{\,\prime}$, denoted by $c\succ^V_{\,rat} c^{\,\prime}$, if $c\succsim^V_{\,rat} c^{\,\prime}$ and $\neg\left(c^{\,\prime} \succsim^V_{\,rat} c\right).$
The author  also characterizes the class of rationality indices consistent with the Violation Criterion, i.e., those scores $I(c)$ assuming values in $\mathbb{R}_+$   for which  $c\succ^V_{\,rat} c^{\,\prime}$ implies $I(c)>I(c^{\,\prime}).$ 
Here I show that the compromise-based degree of irrationality is not consistent with the Violation Criterion.
To see why, consider the choices $c\colon \X\to X$  and $c^{\,\prime}\colon \X\to X$ on $X=\{w,x,y,z\}$, with $c$ defined by

 	$$\underline{w}xyz,\;\;\;w\underline{x}y,\;\;\;\underline{w}xz,\;\;\;\underline{w}yz,\;\;\;\underline{x}yz,\;\;\;\underline{w}x,\;\;\;\underline{w}y,\;\;\;\underline{w}z,\;\;\;\underline{x}y,\;\;\;\underline{x}z,\;\;\;\underline{y}z,$$

and $c^{\,\prime}$ such that 
 	
 	$$\underline{w}xyz,\;\;\;w\underline{x}y,\;\;\;w\underline{x}z,\;\;\;w\underline{y}z,\;\;\;\underline{x}yz,\;\;\;\underline{w}x,\;\;\;\underline{w}y,\;\;\;\underline{w}z,\;\;\;\underline{x}y,\;\;\;\underline{x}z,\;\;\;\underline{y}z.$$
 	
The datasets $c$ and $c^{\,\prime}$ violate WARP under constant selection, since  $w$ meets, for both choices, the requirements of Definition \ref{DEF:violation_alpha_constant_selection}, and, as a consequence, $irr_{\mathsf{Comp}}(c)=irr_{\mathsf{Comp}}(c^{\,\prime})=1.$
However, the reader can check that $c\succ ^{V}_{rat} c^{\,\prime}.$\footnote{To verify this fact, note that, given any $\mathscr{M}\subseteq \mathscr{X}$ containing the menu $wxy$ and at least one between $wxyz$ and $wx$, the partial choice $c_\mathscr{M}$ is not rationalizable, and the same holds true for $c^{\,\prime}_{\mathscr{M}},$ since $c(wxyz)=c^{\,\prime}(wxyz)$, $c(wxy)=c^{\,\prime}(wxy)$, and $c(wx)=c^{\,\prime}(wx)$.
Moreover, for any $\mathscr{M}\subseteq \mathscr{X}$ not including $wxy$, or including $wxy$ and none of the menus $wxyz$ and $wxy$, $c_{\mathscr{M}}$ is always rationalizable.
Thus, I conclude that $c\succsim^{V}_{rat} c^{\,\prime}$.
Note that, given $\mathscr{M}^*=\X\setminus \{\{wxy\}\},$ $c_{\mathscr{M}^*}$ is rationalizable, whereas the partial choice $c^{\,\prime}_{\mathscr{M}^*}$ is not,
yielding $\neg \left(c^{\,\prime}\succsim^{V}_{rat} c\right).$}

\section{Alternative measures of compromise}\label{SECT:alternatives_models_of_compromise}

I now compare my pattern of compromise with other possible models of mediation between opposite judgments.

\subsection{Menu-invariant compromise}\label{SUBSECT:menu_invariant_compromise}

As already mentioned in Section~\ref{SECTION:model} this measure of choice irrationality {stems} from a multi-self approach, which deals with distinct levels of compromise across menus.
However, the extent of the DM's mediation may be fixed \textit{a priori}, thus affecting only the order of the alternatives contained in each menu.
According to this procedure, fixed a preference $\rhd,$ a $i\in\{0,\cdots,\vert X\vert-1\},$ and a menu $A\in\X,$ the DM's first restricts $\rhd$ to the ranking $\rhd^{A}=\{(x,y)\,\text{s.t.}\,(x,y)\in \rhd\,\text{and}\,x,y\in A  \}$ of alternatives contained in $A,$ and then selects from $A$ the option $\max(\rhd^{A}_i,A)$, maximizing the $i$-th compromise on $\rhd^{A}.$\footnote{I would like to thank an anonymous referee for suggesting this variation on my approach.}
This method is represented below.

\begin{definition}\label{DEF:menu_invariant_compromise}
	A choice $c\colon \X\to X$ is with \textit{menu-invariant compromise of extent} $i$ if there is $\rhd\in\mathsf{LO}(X)$ and $i\in\{0,\cdots,\vert X\vert-1\}$ such that $c(A)=\max(A,\rhd^{A}_i)$ for any $A\in \X.$
	I say that $(\rhd,i)$ \textit{is a rationalization} \textit{by menu-invariant compromise of} $c$.  \end{definition} 
	
Menu-invariant compromise of extent $i$ is equivalent to a procedure in which the DM from each menu discards the first $i$ items, and she picks the best option among those available, if any. Otherwise, she chooses the worst alternative.

\begin{definition}\label{DEF:satisficing_model}
	A { choice} $c\colon \X\to X$ is \textit{satisficing with threshold} $i+1$ if there is $\rhd\in\mathsf{LO}(X)$ and $i\in\{0,\cdots,\vert X\vert-1\}$ such that, for any $A\in\X,$ $c(A)=\max(A\setminus A^{\rhd}_i,\rhd)$ if $A\setminus A^{\rhd}_i\neq \es$ (i.e. if $\vert A\vert\geq i+1$), and $c(A)=\min(A,\rhd)$ otherwise.
	 I say that $(\rhd,i)$ is a \textit{rationalization} \textit{by {  menu-invariant}  satisfaction of} $c$.
\end{definition}
	
Satisficing choices recall the procedure originally proposed by \cite{Simon1955}, and later developed, among the others, by \cite{Papi2012} and \cite{BarberadeClippelNemeRozen2022} in revealed preference domains, in  which the DM selects the alternative(s) reaching a satisfaction threshold.
In my framework the chosen option is, if available, \textit{exactly} the one holding in the menu the $(i+1)$-th position with respect to the preference $\rhd.$
{  Note also that any rationalization by {  menu-invariant}  satisfaction of a choice is also a rationalization by satisfaction (as defined in Remark~\ref{REM:menu_dependent_compromise}) of it, in which the index $i$ is fixed, and it does not change across menus.} 
Moreover, the second best procedures discussed in Example~\ref{EXMP:second_best_procedures} of Subsection~\ref{SECTION:compromise_based_degree_irrationality} are choices satisficing with threshold $2$.
As mentioned before, I have:

\begin{lemma}\label{LEM:equivalence_menu_invariant_compromise_satisficing}
Let $c\colon \X\to X$ be a choice on $X$.
Fix some $\rhd\in\mathsf{LO}(X)$ and $i\in\{0,\cdots,\vert X\vert-1\}.$
The pair $(\rhd,i)$ is a rationalization by { menu-invariant} compromise of $c$ if and only if $(\rhd,i)$ is a rationalization by {  menu-invariant}  satisfaction of $c$.
\footnote{The proof of this result is straightforward.
Thus, I leave it to the reader.}
\end{lemma}

Menu-invariant compromise can be inferred from some properties of the dataset. 

\begin{definition}\label{DEF:P_i}
Given a choice $c\colon \X\to X$ on a set $X$, and $i\in\{0,\cdots,\vert X\vert-1\}$, let $P_i$ be the binary relation defined by $xP_i y$ if there is $A\in\X$ s.t. $\vert A\vert\leq i+1,$ $x,y\in A,$ and $y=c(A).$ 
For each $x\in X$ I denote by $M_{\,x}^{P_i^{\uparrow}}$ the set $\{y\in X\colon y\,P_i\,x\}.$
I say that $P_i$ \textit{satisfies $i$-rejection} if, for any $B\in\X$ s.t. $\vert B\vert>i+1,$ I have that $\left\vert M_{\,{ c(B)}}^{P_i^{\uparrow}}\cap B\right\vert=i.$
\end{definition} 

According to Definition~\ref{DEF:P_i} an item $x$ is revealed to be preferred to $y$ if it is offered in a menu, of size lower or equal than $i+1$, from which $y$ is selected.
The set $M_{\,x}^{P_i^{\uparrow}}$ contains all the options superior to $x$, according to the ranking inferred from data.
Finally, $i$-rejection ensures that, whenever an alternative is picked from a menu  of cardinality greater than $i+1,$ then there are also exactly $i$ alternatives better, following the elicited preference, than the chosen one.
When $i>0$, the absence of cycles of any length in $P_i$ and $i$-rejection characterize menu-invariant compromise.
  
\begin{theorem}\label{THM:menu_invariant_characterization}
	Consider a choice $c\colon\X \to X$ and $i\in\{1,\cdots,\vert X\vert-1\}.$
	We have that $c$ is with menu-invariant compromise of extent $i$ if and only { if} $P_i$ is asymmetric, acyclic, and it satisfies $i$-rejection.
	Moreover, if $(\rhd,i)$ is a rationalization by menu-invariant compromise of $c$, then $\rhd\equiv P_i.$ 
\end{theorem}

This result implies that the classes of rationalizable choices, choices with menu-invariant compromise of extent $0,$ and choices with menu-invariant compromise of extent $n-1,$ where $n$ is the cardinality of the ground set, are the same.
Moreover, Theorem~\ref{THM:menu_invariant_characterization} shows also that $P_i$ is the unique preference explaining the observed choice by menu-invariant compromise with extent $i$.
It is worth pointing out that the measure of compromise that can be determined by the model proposed in Definition~\ref{DEF:menu_invariant_compromise} is less flexible than the compromise-based degree of irrationality, since it assumes that the DM  neglects the same number of alternatives on top of each menu.
Moreover, it considers a \textit{naive} subject, who has no awareness of the alternatives that may be provided before the menu is presented to him.
Thus, a mediation between her preference and its negation modifies only the ranking of the available options.  
{As expected, the empirical content of the classes of choices determined by these two indices is different.
Indeed, it can be {shown} that there is a choice whose compromise-based degree of irrationality is $i$, with $0\leq i\leq\vert X\vert-1$, which is not a choice with menu-invariant compromise of extent $i$, and vice versa.\footnote{A formal proof is available upon request.}
}

\subsection{"Top-2" rationality}
\label{SUBSECT:top-2_rationality}          

Another algorithm providing a criterion that partially supports the negation of the DM's preference is the one that swaps the first two alternatives on top. 

\begin{definition}\label{DEF:top_2_rationality}
Given a set $X,$ and $\rhd\in\mathsf{LO}(X),$ denote by $\rhd^{2 }$ the linear order on $X$ such that 
\begin{itemize}
	\item[(i)] for any $a\in X$ and $b\in X\setminus X^{\rhd}_2$, $a\rhd b$ implies $a\rhd^{2 } b,$  and 
	\item [({ ii})] for $a,b\in X^{\rhd}_2$, $a\rhd b$ implies $b\rhd^{2 } a.$
\end{itemize}
		A choice $c\colon \X\to X$ on $X$  is \textit{top-2 rationalizable} if there is $\rhd\in\mathsf{LO}(X)$ such that, for any $A\in\X$ either $c(A)=\max(A,\rhd)$ or $c(A)=\max(A,\rhd^{2 })$ holds.	
\end{definition}

A top-2 rationalizable choice is justified either by the maximization of the DM's preference or by the distortion of it in which the rank of the best two alternatives is inverted.
According to the narrative described in Definition~\ref{DEF:top_2_rationality},  the subject regards in some occasions also the denial of her preference, only partially downgrading her best option.
This pattern resembles second best procedures, in which the DM always chooses, among the alternatives available, the option that immediately follows the best one.
However, differently from them, it allows the picked alternative from some menus to be the one on top of the DM's preference.
Moreover, the swap between the best and the second best alternative modifies the DM's evaluation of all the alternatives belonging to the ground set, not only those contained in the menus.   
Thus, top-2 rationalizable choice behavior brings a distinct empirical content, and it can be inferred from the following axiom.

\begin{definition}\label{DEF:WARP_exception_two_options}
A choice $c\colon\X\to X$ satisfies \textit{WARP with the exception of two items on top} if there are unique and distinct $x,y\in X$ such that
\begin{itemize}  
	\item[(i)] if there are $A,B\in\X$ such that $(A,B)$ is a reversal, then either $c(A)=x,c(B)=y$ or $c(A)=y,c(B)=x$ holds, and 
	\item[(ii)] for any $D\in \X$ such that $D\cap xy\neq \es$, I have that $c(D)\in D\cap xy$. 
\end{itemize}	
\end{definition} 
 
WARP with the exception of two items on top  requires that failures of WARP happen only if in an observed reversal the two chosen alternatives are always the same.
Moreover, these alternative{ s} are always chosen, if offered.
This axiom characterizes top-2 rationality. 
 
 \begin{theorem}\label{THM:top_2_rationality_characterized_warp_ with the exception of two items.}
 	A choice is top-2 rationalizable if and only if it satisfies WARP with the exception of two items on top.
 \end{theorem}
 
{  Moreover, there are only two preferences that explain a top-2 rationalizable choice.
Let $c\colon \X\to X$ be a choice satisfying WARP with the exception of two items on top, and let $x,y\in X$ be the two alternatives such that conditions (i) and (ii) of Definition~\ref{DEF:WARP_exception_two_options} are satisfied.
	Let $R_1,R_2$ be the binary relations defined, for any $i\in\{1,2\}$, by $w\, R_{i}\, z$ for any $w,z\in X$ such that $\{w,z\}\neq \{x,y\},$ and there is { $B\in\X$} such that $z\in B$ and $w=c(B).$
 Moreover, set $x\,R_1\,y$, $\neg(y\,R_1\,x)$ and $y\,R_2\,x$, $\neg(x\,R_2\,y).$ 
 The following result holds.}

\begin{lemma}\label{LEM:uniqueness_preference_top_2_rationality}
	Assume that $c\colon \X\to X$ satisfies  WARP with the exception of two items on top.
	Then $R_1,R_2$ are linear orders, and for any $\rhd^{\prime}\in\mathsf{LO}(X)$ such that $c(A)=\max(A,\rhd^{\prime})$ or $c(A)=\max(A,\rhd^{\prime\,2 })$ holds {  for any $A\in\X$}, I have that either $\rhd^{\prime}=R_1$ or $\rhd^{\prime}=R_2$ is true.    
\end{lemma}

Although it is based on a plausible economic interpretation, the procedure discussed in this subsection describes a compromise on the DM's taste of \textit{limited magnitude}, which affects only the order of the two best alternatives.
Thus, it does not allow for a general assessment of the degree of rationality of a dataset.
{  Rationalizations by compromise, on the other hand, evaluate all levels of mediation between opposing criteria, yielding a measure of irrationality of the choice.
A potential extension of top-2 rationality  would explain the selection from each menu by reversing the order of the first $i$ alternatives of the ground set, while leaving the rest unchanged: this paradigm is discussed in the next subsection.
}

\subsection{Nudging}
\label{SUBSECT:nudging}

{  In Section~\ref{SECTION:model} and Subsection~\ref{SECT:alternatives_models_of_compromise}\ref{SUBSECT:menu_invariant_compromise} I assumed that compromise  moves the first $i$ alternatives, contained either in the ground set or in some menu, to the bottom of the DM's ranking.
However, one can consider a different modification of individual tastes, in which the DM,
to favor the opposite judgment, may invert the order of some items on top of her preference.
}

{  \begin{definition}\label{DEF:nudging}
Given a set $X$, some $\rhd\in\mathsf{LO}(X)$, and $1\leq i \leq \vert X\vert,$ the \textit{i-th nudging on $\rhd$} is the binary relation, denoted by $\rhd^{i}$, such that
\begin{enumerate}[\rm(i)]
	
	\item for any $a\in X$ and $b\in X\setminus X^{\rhd}_i$, $a\rhd b$ implies $ a\rhd^{i} b$, and
	\item for any $a,b\in X^{\rhd}_i$, $a\rhd b$ implies $b\rhd^i a.$ 
\end{enumerate} 
A linear order $\rhd^{\prime}\in\mathsf{LO}(X)$ is \textit{a nudging on $\rhd$} if $\rhd^{\prime}\equiv \rhd^i $ for some $i\in\{1,\cdots,\vert X\vert\}.$
 I denote by $\mathsf{Nudg}(\rhd)$ the family $\{\rhd_i\}_{1\leq\, i\,\leq \vert X\vert}$ of all the $\vert X\vert$ nudging on $\rhd$.
 Let  $c\colon \X \to X$ be a choice on a set $X$, and $\rhd$ be a linear order in $\mathsf{LO}(X)$.
			A family $\mathsf{Nudg}_c(\rhd)\subseteq \mathsf{Nudg}(\rhd)$ is a \textit{rationalization by nudging on $\rhd$ of $c$} if for every $A\in\X$ there is $\rhd^i\in\mathsf{Nudg}_c(\rhd)$ such that $c(A)=\max(A,\rhd^i)$.
		Finally let
 $$irr_{\mathsf{Nudg}}(c)=\min_{\rhd\in\mathsf{LO}(X)}\left(\min_{\mathsf{Nudg}_c(\rhd)\subseteq \mathsf{Nudg(\rhd)}}\left(\max_{i\colon\rhd^i\in\mathsf{Nudg}_c(\rhd)} i\right)\right)$$
	the \textsl{nudging-based degree of irrationality} of $c$.
\end{definition}}

{  In a nudging on her preference, the DM reverses the order of the first $i$ alternatives of the ground set}.\footnote{  Note that $\rhd^1\equiv \rhd.$}
{  The connection with the model proposed in the previous subsection is evident: a pair $(\rhd,\rhd^{  2})$ explaining a top-2 rationalizable  choice is also a rationalization by nudging  of the same dataset.}
As for compromise, a rationalization by nudging explains the dataset by distinct levels of self-prompting.
Thus, the nudging-based degree of irrationality determines a lower bound to the minimal intensity of nudging needed to justify an observed choice.\footnote{  Differently from the compromise-based degree of irrationality, $irr_{\mathsf{Nudg}}(c)=1$  if $c$ is rationalizable, and $irr_{\mathsf{Nudg}}(c)>1$ otherwise.}
The economic interpretation of this phenomenon is close to the one proposed for compromise: the subject nudges herself to do better, evaluating the first $i$ alternatives according { to} the negation of her preference.
{  However, differently from the distortion of Definition~\ref{DEF:preference_compromise}, the relation between such items and the remaining ones does not change.}
Indeed, each linear order can be generated either by compromise or nudging, appealing to the tension between the same conflicting criteria.

{ \begin{lemma}\label{LEM:compromise_equals_nudging}
	Let $\rhd,\rhd^{\prime}\in\mathsf{LO}(X)$ be linear orders on $X$.
	I have that $\rhd^{\prime}\in\mathsf{LO}(X)$ is a compromise on $\rhd$ if and only if it is a nudging on $-\rhd.$
	Moreover, $\rhd^{\prime}\equiv \rhd_i \equiv -\rhd^{\vert X\vert-i},$ with $i\in\{0,\cdots,\vert X\vert-1\}.$
\end{lemma}

The consequences of mediation and self-prompting on the DM's taste are specular: a compromise on a preference equals a nudging on its negation.  
Thus, these two processes generate the same collection of rankings explaining a choice.

\begin{corollary}

	Let $c\colon \X\to X$ be a choice on $X$.
	I have that $\mathsf{Comp}_c(\rhd)$ is rationalization by compromise on $\rhd$ of $c$ if and only if it is a rationalization by nudging on $-\rhd$ of $c$.
\end{corollary}

As expected, the compromise-based degree of irrationality and the nudging-based degree of irrationality offer distinct predictions, and sort choices in different ways, as shown in the next example.

\begin{example}\label{EXMP:compromise_nudging_degree}

Consider the choices $c\colon \X\to X$ and $c^{\,\prime}\colon \X\to X$ respectively defined by

$$w\underline{x}yz,\;\; \underline{w}xy,\;\; w\underline{x}z,\;\;w\underline{y}z,\;\; \underline{x}yz, \;\; \underline{w}x,\;\; \underline{w}y,\;\; w\underline{z},\;\;\underline{x}y, \;\; \underline{x}z, \;\; \underline{y}z,$$	

and 

$$wx\underline{y}z,\;\; w\underline{x}y,\;\; w\underline{x}z,\;\;\underline{w}yz,\;\; \underline{x}yz, \;\; \underline{w}x,\;\; \underline{w}y,\;\; \underline{w}z,\;\;\underline{x}y, \;\; \underline{x}z, \;\; \underline{y}z.$$

The reader can check that $irr_{\mathsf{Comp}}(c)=1$, since $c$ is not rationalizable, and the pair $(\rhd,\rhd_1)$, with $\rhd$ defined by $w\rhd x\rhd y\rhd z$, is a rationalization by compromise on $\rhd$ of $c$.
Moreover, $irr_{\mathsf{Comp}}(c^{\,\prime})=2$, since $c^{\,\prime}$ violates WARP under constant nonreciprocal selection of $2$ items, namely $w$ and $x$.
However, $irr_{\mathsf{Nudg}}(c)=4$, because any option is selected at least once, requiring a complete  turnaround of any preference, whereas $irr_{\mathsf{Nudg}}(c^{\,\prime})=2$, since $c^{\,\prime}$ is not rationalizable, and $(\rhd,\rhd^{1})$ is a rationalization by nudging on $\rhd$ of $c^{\,\prime}$.
Thus $irr_{\mathsf{Comp}}(c)<irr_{\mathsf{Comp}}(c^{\,\prime})$, whereas $irr_{\mathsf{Nudg}}(c)>irr_{\mathsf{Nudg}}(c^{\,\prime}),$ suggesting that these two indices measure the rationality of a choice in light of different features.
\end{example}}

\section{Concluding remarks}\label{SECT:concluding_remarks}

In this work I assume that violations of rationality in choices stem from the DM's tension opposite judgments.  
When presented with a menu, the DM adopts some compromise on her true preference, in which some of the best alternatives are moved, in reverse order, to the bottom.
This paradigm allows to classify observed choices according to a measure of irrationality, consisting of a lower bound to the maximal index of the compromises that the DM applied in her decision.
I single out the necessary and sufficient conditions to estimate the compromise-based degree of irrationality of a choice, and I partially elicit the linear orders whose compromises are needed to justify the DM's picks.
Non rationalizable choices displaying the lowest degree of compromise are characterized, and the compromises explaining the datasets are identified.
I provide an alternative and manageable characterization of choices that exhibit the highest degree of compromise, and I prove their ubiquity.
Finally alternative measurements of compromise are investigated, and compared with the compromise-based degree of irrationality.

In such framework compromise is menu-dependent, but there is no mechanism that matches menus and the maximizing compromises on the DM's true preference.
However, the trade-off between conflicting criteria determine well-known phenomena, such as temptation, reciprocity, guilt, self-punishment, and self-confidence, in which the framing of the DM's choice is crucial.
Thus, future research may be devoted to propose a rationality measure which takes into account \textit{motives} of the DM's mediation, by formally defining some rule that associates compromises to menus.
Moreover, a natural extension of my setting may consider some \textit{randomization} between the DM's compromise to rationalize and sort \textit{stochastic choices}, i.e. dataset{ s} displaying, for each menu, the probability of selecting some item.
Consequently, stochastic compromise may become a subclass of the \textit{Random Utility Models} (\textit{RUMs}) \citep{BlockMarschak1960} with support limited to the compromises on the DM's true preference.\footnote{This method is employed in \cite{Petralia2025}.}
Finally, a further promising (but non-trivial) extension of this work may lead to a{ n} index of irrationality for \textit{stochastic intertemporal choices} \citep{FudengbergIijimaStrzalecki2015,LuSaito2018}, assuming the DM's current compromise depends also on her past one. 
 
\section*{Appendix: Proofs}

\noindent {\textbf{\large Proof of Lemma~\ref{LEMMA:RSP_non_testable_method}}.
Let $c\colon\X\to X$ be a choice.
	Given $\rhd\in\mathsf{LO}(X)$ and $x\in X$, denote by $x^{\uparrow\rhd}$ the set $\{y\in X\colon y\rhd x\}$.
	Observe that $c(A)=\max\left(A,\rhd_{\left\vert  c(A)^{\uparrow\rhd}\right\vert}\right)$, and $0\leq \left\vert c(A)^{\uparrow\rhd} \right\vert \leq \vert X\vert -1$ for any $A\in\X$. 
	Thus, the family $\left\{\rhd_{\left\vert c(A)^{\uparrow\rhd} \right \vert}\right\}_{A\in\X}$ is a rationalization by compromise on $\rhd$ of $c$. 
	\qed

\smallskip

\noindent \textbf{\large Proof of Theorem~\ref{THM:characterization_compromise_based_degree_irrationality}.}
I need some preliminary notation and results.
Given $\rhd\in\mathsf{LO}(X)$, order the ground set $X$ as $\left\{x^{\rhd}_1,\cdots,x^{\rhd}_{\vert X\vert}\right\},$ where $x^{\rhd}_i\rhd x^{\rhd}_j$ if and only if $i<j.$
Thus, given some $1\leq h\leq\vert X\vert,$ $x^{\rhd}_{h}$ denotes the item of $X$ holding the $h$-th position in $X$ with respect to $\rhd.$ 
Denote by $x_h^{\uparrow\rhd}$ the set $\{y\in X \colon y\rhd x^{\rhd}_h \}$, and by $x_h^{\downarrow\rhd}$ the set $\{y\in X \colon x^{\rhd}_h\rhd y \}.$
Finally, given $h\in\{1,\cdots,\vert X\vert\}$, denote by $\mathscr{C}_{x^{\rhd}_{h}}$ the set $\bigcup_{A\in\X\colon x^{\rhd}_{h}=c(A)}A$. I have:

\begin{lemma}\label{LEM:necessary_condition_compromises}
Given $\rhd\in\mathsf{LO}(X)$, for any $g\in\{0,\cdots,\vert X\vert-1\}$ and any $h,i\in\{1,\cdots\vert X\vert\}$ such that $g<h<i$, I have that $x^{\rhd}_{h}\rhd_g x^{\rhd}_{i}.$ 
\end{lemma}

\begin{proof}
It follows from Definition~\ref{DEF:preference_compromise}.
\end{proof}

\begin{lemma}\label{LEM:no_external_switches}
	Let $c\colon\X\to X$ be a choice on $X$.
		Given $\rhd\in\mathsf{LO}(X)$, let $\mathsf{Comp}_{c}(\rhd)$ be a rationalization by compromise on $\rhd$ of $c$ such that $\max_{i\colon\rhd_i\in\mathsf{Comp}_c(\rhd)}i=j,$ for some $1\leq j\leq \vert X\vert-1.$
Then, there is no reversal $(A,B)$ such that $c(A)\neq x^{\rhd}_{h}\neq c(B)$ for any $h\in\{0,\cdots,j\}$.
\end{lemma}

\begin{proof}
	Toward a contradiction, assume that $\mathsf{Comp}_{c}(\rhd)$ is a rationalization by compromise on $\rhd$ of $c$ such that $\max_{\,i\colon\rhd_i\in\mathsf{Comp}_c(\rhd)}i=j,$ for some $0\leq j\leq \vert X\vert-1,$ and there is a reversal $(A,B)$ such that $c(A)\neq x^{\rhd}_{h}\neq c(B)$ for any $h\in\{0,\cdots,j\}.$
	Thus, by Definition~\ref{DEF:minimal_violations_of_alpha} there are menus $A,B\in\X$ and $k,l\in\{j+1,\cdots,\vert X\vert\}$ such that $k<l$, $x^{\rhd}_{k},x^{\rhd}_{l}\in (A\cap B)$, and either $x^{\rhd}_{k}=c(B)$, and $x^{\rhd}_{l}=c(A)$, or $x^{\rhd}_{l}=c(B)$ and $x^{\rhd}_{k}=c(A)$ hold.
	{  Definition~\ref{DEF:rationalization_by_compromise} implies that there are $f,g\in\{0,\cdots,j\}$  such that $x^{\rhd}_k\rhd_f x^{\rhd}_l$ and $x^{\rhd}_l \rhd_g x^{\rhd}_k,$ which contradicts Lemma~\ref{LEM:necessary_condition_compromises}.}
\end{proof}

\begin{lemma}\label{LEM:self_punishment_j_successors_empty}
	{  Let $c\colon\X\to X$ be a choice on $X$.
	Given $\rhd\in\mathsf{LO}(X)$, let $\mathsf{Comp}_{c}(\rhd)$ be a rationalization by compromise on $\rhd$ of $c$ such that $\max_{i\colon\rhd_i\in\mathsf{Comp}_c(\rhd)}i=j,$ for some $1\leq j\leq \vert X\vert-1.$
	Assume now that $c(A)=x^{\rhd}_l$, for some $j< l\leq \vert X\vert.$ 
	Then, I have that
	$A\,\cap\left(x^{\uparrow\rhd}_{j}\cap x^{\downarrow\rhd}_{l}\right)=\es$.}
\end{lemma}

\begin{proof}
{  Assume toward a contradiction that there is some $x^{\rhd}_{k}\in A$, with $j<k<l.$
	Definition~\ref{DEF:preference_compromise} and $\max_{i\colon\rhd_i\in\mathsf{Comp}_c(\rhd)}i=j$ imply that there is no  $d\in\{0,\cdots,j\}$ and $\rhd_{d}\in\mathsf{Comp}_{c}(\rhd)$ such that $x^{\rhd}_{l}=\max(A,\rhd_d),$ a contradiction. } 
	\end{proof}

\begin{lemma}\label{LEM:selection_reversal_respects_alpha}
Let $c\colon\X\to X$ be a choice on $X$.
	Given $\rhd\in\mathsf{LO}(X)$, let $\mathsf{Comp}_{c}(\rhd)$ be a rationalization by compromise on $\rhd$ of $c$ such that $\max_{i\colon\rhd_i\in\mathsf{Comp}_c(\rhd)}i=j$ holds, for some $1\leq j\leq \vert X\vert-1.$
	Given $h\in\{1,\cdots,j\},$ denote by $l$ the index in $\{j+1,\cdots,\vert X\vert\}$ such that {  $x^{\rhd}_l=\max(\mathscr{C}_{x^{\rhd}_{h}}\cap x^{\uparrow \rhd}_j,\rhd).$}
	If there is no reversal $(A,B)$ such that either $x_h=c(A),y=c(B),$ or $y=c(A),x_h=c(B)$ hold, with $y\in X\setminus \{x^{\rhd}_1,\cdots,x^{\rhd}_{j}\}$, then for any $D\in \X$ such that $x^{\rhd}_h,x^{\rhd}_m\in D,$ with $l\leq m\leq \vert X\vert$, I have that $c(D)\neq x^{\rhd}_m.$ 
\end{lemma}

\begin{proof}
Toward a contradiction, assume that there is $l\leq m$ and some $D\in\X$ such that $x^{\rhd}_h,x^{\rhd}_m\in D,$ and $c(D)=x^{\rhd}_m.$
Denote by $E$ a set such that $x^{\rhd}_h,x^{\rhd}_l\in E,$ and $c(E)=x^{\rhd}_h.$
If $m=l,$ the pair $(D,E)$ is a reversal in which $x^{\rhd}_{h}$ and $x^{\rhd}_l$, with $h<l$, are selected, and I get a contradiction.
Thus, assume $l<m\leq \vert X\vert-1.$
Note that, since $\mathsf{Comp}_{c}(\rhd)$ is a rationalization by { compromise} of $c$ by $\rhd$, $\max_{i\colon\rhd_i\in\mathsf{Comp}_c(\rhd)}i=j,$ and $j<l<m$, Definition~\ref{DEF:rationalization_by_compromise}  implies that $c(x^{\rhd}_l,x^{\rhd}_m)=x^{\rhd}_l.$
Consider the menu $x^{\rhd}_h x^{\rhd}_{j+1} x^{\rhd}_m.$
If $c(x^{\rhd}_h x^{\rhd}_{j+1} x^{\rhd}_m)=x^{\rhd}_h,$ then the pair $(x^{\rhd}_h x^{\rhd}_{j+1} x^{\rhd}_m,D)$ is a reversal in which $x^{\rhd}_h$ and $x^{\rhd}_m$, with $m>h$ are selected, a contradiction.
If 	$c(x^{\rhd}_h x^{\rhd}_l x^{\rhd}_m)=x^{\rhd}_m,$ then the pair $(x^{\rhd}_h x^{\rhd}_l x^{\rhd}_m,x^{\rhd}_l x^{\rhd}_m)$ is a reversal, which is impossible by Lemma~\ref{LEM:no_external_switches}.
Finally, if 	$c(x^{\rhd}_h x^{\rhd}_l x^{\rhd}_m)=x^{\rhd}_l,$ then, the pair $(D,E)$ is a reversal in which $x^{\rhd}_h$ and $x^{\rhd}_l$, with $h<l$, are selected, a contradiction.
We conclude that $c(D)\neq x^{\rhd}_m.$
 \end{proof}

\begin{lemma}\label{LEM:switches_needed_in_the_first_j_positions}
Let $c\colon\X\to X$ be a choice on $X$.
{  Assume that $irr_{\mathsf{Comp}}(c)=j$ and let $\rhd\in\mathsf{LO}(X)$ be a linear order such that $\mathsf{Comp}_{c}(\rhd)$ is rationalization by compromise on $\rhd$ of $c$ and $\max_{\,i\colon\rhd_i\in\mathsf{Comp}_c(\rhd)}i=j,$ with $1\leq j\leq \vert X\vert-1$.}
Then for { each} $h\in\{1,\cdots,j\},$  there is a reversal $(A,B)$ such that either $c(A)=x^{\rhd}_h$ and $c(B)=y$ or $c(A)=y$ and $c(B)=x^{\rhd}_h,$ with $y\in X\setminus\{x^{\rhd}_1,\cdots,x^{\rhd}_j\}.$
\end{lemma}

\begin{proof}
Assume toward a contradiction that there is $h\in\{1,\cdots,j\}$ such that there is no reversal $(A,B)$ for which either $c(A)=x^{\rhd}_h$ and $c(B)=y$ or $c(A)=y$ and $c(B)=x^{\rhd}_h$ holds, with $y\in X\setminus\{x^{\rhd}_1,\cdots,x^{\rhd}_j\}.$

{  If $\mathscr{C}_{x^{\rhd}_{h}}\cap x^{\uparrow\rhd}_{j}=\es,$} then consider the linear order $\rhd^{\prime}$ defined by $x^{\rhd^{\prime}}_{g}=x^{\rhd}_{g}$ for any $1\leq g<h$, $x^{\rhd^{\prime}}_{g}=x^{\rhd}_{g+1}$ for any $h\leq g< \vert X\vert,$ and $x^{\rhd^{\prime}}_{\vert X\vert}=x^{\rhd}_{h}.$   
 The linear order $\rhd^{\prime}$ is obtained from $\rhd$ by moving the $h$-th item to the bottom, and leaving the rank of the other alternatives unchanged.
Let $\{\rhd^{\prime},\rhd^{\prime}_1,\cdots,\rhd^{\prime}_{j-1}\}$ the set of compromises on $\rhd^{\prime}$ up to the index $j-1$.
I show that $\{\rhd^{\prime},\rhd^{\prime}_1,\cdots,\rhd^{\prime}_{j-1}\}$ is a rationalization by compromise on $\rhd^{\prime}$ of $c$ .
Consider a menu $A\in\X.$
If $c(A)=x^{\rhd^{\prime}}_g$ for a $g\in \{1,\cdots,\vert X\vert\}$, and  $A\cap \{x^{\rhd^{\prime}}_{1},\cdots,x^{\rhd^{\prime}}_{g-1}\}=\es,$ then the definition of $\rhd^{\prime}$ and Definition~\ref{DEF:preference_compromise} imply that $c(A)=\max(A,\rhd^{\prime})$.
If $c(A)=x^{\rhd^{\prime}}_g$, for some $g\in \{1,\cdots,j\}$, and $A\cap \{x^{\rhd^{\prime}}_{1},\cdots,x^{\rhd^{\prime}}_{g-1}\}\neq\es,$ then the definition of $\rhd^{\prime}$ and Definition~\ref{DEF:preference_compromise} $c(A)=\max(A,\rhd^{\prime}_{g-1}).$
Assume now that $c(A)=x^{\rhd^{\prime}}_g$, for some $j<g< \vert X\vert,$ and $A\cap \{x^{\rhd^{\prime}}_{1},\cdots,x^{\rhd^{\prime}}_{g-1}\}\neq\es.$
{  Note that the definition of $\rhd^{\prime}$ and  Lemma~\ref{LEM:self_punishment_j_successors_empty}  imply that $A\cap\left(x^{\uparrow\rhd^{\prime}}_{j-1}\cap x^{\downarrow\rhd^{\prime}}_{g}\right)=A\cap\left(x^{\uparrow\rhd}_{j}\cap x^{\downarrow\rhd}_{g+1}\right)=\es.$}
Thus, Definition~\ref{DEF:preference_compromise} and the definition of $\rhd^{\prime}$ yield that $c(A)=\max(A,\rhd^{\prime}_{j-1})$.
Assume now that $c(A)=x^{\rhd^{\prime}}_{\vert X\vert}$ and $A\cap \{x^{\rhd^{\prime\prime}}_{1},\cdots,x^{\rhd^{\prime}}_{\vert X\vert-1}\}\neq\es.$
Since $\mathscr{C}_{x^{\rhd}_{h}}\cap x^{\downarrow\rhd}_{j}=\es,$ the definition of $\rhd^{\prime}$ and Definition~\ref{DEF:preference_compromise} yield $c(A)=\max(A,\rhd^{\prime}_{j-1}).$
{  Since $c(A)=\max(A,\rhd^{\prime}_i)$ for some $i\in\{1,\cdots,j-1\},$ Definition~\ref{DEF:rationalization_by_compromise} implies that $\{\rhd^{\prime},\rhd^{\prime}_1,\cdots,\rhd^{\prime}_{j-1}\}$ is a rationalization by compromise on $\rhd^{\prime}$ of $c$. }
Since $\{\rhd^{\prime\prime},\rhd^{\prime}_1,\cdots,\rhd^{\prime}_{j-1}\}$ is a rationalization by compromise on $\rhd^{\prime}$ of $c$, and $\max_{i\colon\rhd_i\in\mathsf{Comp}_c(\rhd^{\prime})}i=j-1,$ I conclude that $irr_{\mathsf{Comp}}(c)<j,$ a contradiction. 

{  If $\mathscr{C}_{x^{\rhd}_{h}}\cap x^{\uparrow\rhd}_{j}\neq\es,$ }denote by $k$ the index such that {  $x^{\rhd}_{k}=\max(\mathscr{C}_{x^{\rhd}_{h}}\cap x^{\uparrow\rhd}_{j},\rhd).$}
Let $\rhd^{\prime\prime}$ be the linear order defined by $x^{\rhd^{\prime\prime}}_g=x^{\rhd}_g$ for any $1\leq g< h,$ and $k\leq g\leq \vert X\vert$, $x^{\rhd^{\prime\prime}}_g=x^{\rhd}_{g+1}$ for any $h\leq g <k-1,$ and $x^{\rhd^{\prime\prime}}_{k-1}=x^{\rhd}_h.$
The linear order $\rhd^{\prime\prime}$ is obtained from $\rhd$ by moving the $h$-th item immediately before the $k$-th item, without altering the rank of the remaining options.
I show that $\{\rhd^{\prime\prime},\rhd^{\prime\prime}_1,\cdots,\rhd^{\prime\prime}_{j-1}\}$ is a rationalization by self-punishment of $c$ by $\rhd^{\prime\prime}$.
Consider a menu $A\in\X.$
If $c(A)=x^{\rhd^{\prime\prime}}_g$, for some $g\in \{1,\cdots,\vert X\vert\}$, and $A\cap \{x^{\rhd^{\prime\prime}}_{1},\cdots,x^{\rhd^{\prime\prime}}_{g-1}\}=\es,$ then the definition of $\rhd^{\prime\prime}$ and Definition~\ref{DEF:preference_compromise} yield $c(A)=\max(A,\rhd^{\prime\prime})$. 
If $c(A)=x^{\rhd^{\prime\prime}}_g$ for some $g\in \{1,\cdots,j\}$, and $A\cap \{x^{\rhd^{\prime\prime}}_{1},\cdots,x^{\rhd^{\prime\prime}}_{g-1}\}\neq\es,$ then the definition of $\rhd^{\prime\prime}$ and Definition~\ref{DEF:preference_compromise} imply $c(A)=\max(A,\rhd^{\prime\prime}_{g-1}).$
Assume $c(A)=x^{\rhd^{\prime\prime}}_g$  for some $g\in \{j+1,\cdots,k-1\},$ and $A\cap \{x^{\rhd^{\prime\prime}}_{1},\cdots,x^{\rhd^{\prime\prime}}_{g-1}\}\neq\es.$
The definition of $\rhd^{\prime\prime}$ and Lemma~\ref{LEM:self_punishment_j_successors_empty}  imply that {  $A\cap\left(x^{\uparrow\rhd^{\prime\prime}}_{j-1}\cap x^{\downarrow\rhd^{\prime\prime}}_{g}\right)=A\cap\left(x^{\uparrow\rhd}_{j}\cap x^{\downarrow\rhd}_{g+1}\right)=\es.$}
Thus, Definition~\ref{DEF:preference_compromise} and the definition of $\rhd^{\prime\prime}$ yield $c(A)=\max(A,\rhd^{\prime\prime\prime}_{j-1}).$ 
If $c(A)=x^{\rhd^{\prime\prime}}_{k-1},$ then the  definition of $\rhd^{\prime\prime}$ and the fact that  $x^{\rhd}_{k}=\max(\mathscr{C}_{x^{\rhd}_{h}}\cap x^{\downarrow\rhd}_{h},\rhd)$ imply that {  $A\cap\left(x^{\uparrow\rhd^{\prime\prime}}_{j-1}\cap x^{\downarrow\rhd^{\prime\prime}}_{k-1}\right)=\es.$}
Thus, the definition of $\rhd^{\prime\prime}$ and Definition~\ref{DEF:preference_compromise} imply that $c(A)=\max(A,\rhd^{\prime\prime}_{j-1}).$
Finally, assume that $c(A)=x^{\rhd^{\prime\prime}}_g$ for some $g\in\{k,\cdots,\vert X\vert\},$ and $A\cap \{x^{\rhd^{\prime\prime}}_{1},\cdots,x^{\rhd^{\prime\prime}}_{g-1}\}\neq\es.$
Since $\mathsf{Comp}_c(\rhd)$ is a rationalization by compromise on $\rhd$ of $c$ and $\max_{i\colon\rhd_i\in\mathsf{Comp}_c(\rhd)}i=j,$ {  Definition~\ref{DEF:rationalization_by_compromise} implies that $A\cap\left(x^{\uparrow\rhd}_{j}\cap x^{\downarrow\rhd}_{g}\right)=\es.$
By Lemma~\ref{LEM:selection_reversal_respects_alpha} I know that $x^{\rhd}_h=x^{\rhd^{\prime\prime}}_{k-1}\not\in A.$}
These two facts and the definition of $\rhd^{\prime\prime}$ imply that {  $A\cap\left(x^{\uparrow\rhd^{\prime\prime}}_{j-1}\cap x^{\downarrow\rhd^{\prime\prime}}_{k}\right)=\es.$ }
Definition \ref{DEF:preference_compromise} and the definition of $\rhd^{\prime\prime\prime}$ yield {  $c(A)=\max(A,\rhd^{\prime\prime}_{j-1}).$}
Since $c(A)=\max(A,\rhd^{\prime\prime}_i)$ for some $i\in\{1,\cdots,j-1\},$ Definition~\ref{DEF:rationalization_by_compromise} implies that $\{\rhd^{\prime\prime},\rhd^{\prime\prime}_1,\cdots,\rhd^{\prime\prime}_{j-1}\}$ is a rationalization by compromise on $\rhd^{\prime\prime}$ of $c$.
Since $\{\rhd^{\prime\prime},\rhd^{\prime\prime}_1,\cdots,\rhd^{\prime\prime}_{j-1}\}$ is a rationalization by compromise on $\rhd^{\prime\prime}$ of $c$, and $\max_{i\colon\rhd_i\in\mathsf{Comp}_c(\rhd^{\prime\prime})}i=j-1,$ I conclude that $irr_{\mathsf{Com}}(c)<j,$ a contradiction.
\end{proof}

({\bf\textit{Only if part}}).
Let $c\colon\X\to X$ be a choice on $X$, and assume that $irr_{\mathsf{Comp}}(c)=j,$ with $1\leq j\leq \vert X\vert-1.$ 
Definition~\ref{DEF:compromise_based_degree_irrationality} implies that there is $\rhd\in\mathsf{LO}(X)$ and $\mathsf{Comp}_c(\rhd)\subseteq \mathsf{Comp}(\rhd)$ such that {  $\max_{\,i\colon\rhd_i\in\mathsf{Comp}_c(\rhd)}i=j,$} and { $\max_{\,i\colon\rhd^{\prime}_i\in\mathsf{Comp}_c(\rhd^{\prime})}i\geq j$} for any $\rhd^{\prime}\in\mathsf{LO}(X)$ and $\mathsf{Comp}_c(\rhd^{\prime})\subseteq \mathsf{Comp}(\rhd^{\prime}).$
 Assume toward a contradiction that $c$ does not violate WARP under constant nonreciprocal selection of $j$ items.
 Thus, by Definition~\ref{DEF:violation_independent_selection_i_items}, at least one of the following conditions hold:
 \begin{enumerate}[\rm(i)]
\item there is a set $D\subset X$ of cardinality $\vert X\vert =f<j$ such that for any reversal $(A,B)$ I have or $c(A)\in D,$ or $c(B)\in D,$ or both;
 \item  for any $\{x_1,\cdots,x_j\}\subset X$  at least one of the conditions below is true:  
 \begin{enumerate}
\item there is a reversal $(A,B)$ such that $c(A)\neq x_h\neq c(B)$ for any $h\in\{1,\cdots,j\},$ and
\item	there is $h\in\{1,\cdots,j\}$ for which there is no reversal $(A,B)$ such that  either $x_h=c(A),y=c(B),$ or $y=c(A),x_h=c(B)$ holds, with $y\in X\setminus \{x_1,\cdots,x_{j}\}$.  
\end{enumerate}
 \end{enumerate}

If (i) holds, first, denote by $x^{\rhd}_{i,D}$ the item holding the $i$-th place in the set $D$  with respect to $\rhd.$
 Let $\rhd^{\prime\prime\prime}$ be the linear order defined by $x^{\rhd^{\prime\prime\prime}}_g=x^{\rhd}_{g,D}$ for any $g\in\{1,\cdots,\vert D\vert\},$ and $x^{\rhd^{\prime\prime\prime}}_{g}=x^{\rhd}_{g+\vert \{x\in D\colon x^{\rhd}_g \rhd\,x\} \vert}$   for any $g\in \{\vert D\vert+1,\cdots\vert X\vert\}.$
The linear order $\rhd^{\prime\prime\prime}$ moves the alternatives in $D$ to the first $\vert D\vert$ positions, respecting their ranking according to $\rhd$, and  preserves the ordering between all  the remaining options.
I show that $\{\rhd^{\prime\prime},\rhd^{\prime\prime\prime}_1,\cdots,\rhd^{\prime\prime\prime}_{\vert D\vert}\}$ is a rationalization by compromise on $\rhd^{\prime\prime\prime}$  of $c$.
Consider a menu $A\in\X.$
If $c(A)=x^{\rhd^{\prime\prime\prime}}_g$ for some $g\in\{1,\cdots,\vert D\vert\},$ then the definition of $\rhd^{\prime\prime\prime}$ and Definition~\ref{DEF:preference_compromise} yields $c(A)=\max(A,\rhd_{g-1}).$
If $c(A)=x^{\rhd^{\prime\prime\prime}}_g$ for some $g\in\{\vert D\vert+1,\cdots,\vert X\vert\},$ and $A\setminus (\{x^{\rhd^{\prime\prime\prime}}_{\vert D+1\vert},\cdots, x^{\rhd^{\prime\prime\prime}}_{\vert X\vert}\}\setminus x^{\rhd^{\prime\prime\prime}}_g)=\es,$ then $c(A)=\max\left(A,\rhd^{\prime\prime\prime}_{\vert D\vert}\right).$
Assume that {  $c(A)=x^{\rhd^{\prime\prime\prime}}_g$ for some $g\in\{\vert D\vert+1,\cdots,\vert X\vert\},$ and $A\setminus (\{x^{\rhd^{\prime\prime\prime}}_{\vert D+1\vert},\cdots, x^{\rhd^{\prime\prime\prime}}_{\vert X\vert}\}\setminus x^{\rhd^{\prime\prime\prime}}_g)\neq \es.$}
Lemma~\ref{LEM:switches_needed_in_the_first_j_positions}  implies that $(A\setminus \{x^{\rhd^{\prime\prime\prime}}_{\vert D+1\vert},\cdots, x^{\rhd^{\prime\prime\prime}}_{\vert X\vert}\})\cap\{x_1^{\rhd},\cdots,x_j^{\rhd}\}=\es,$ and, in particular, that $x^{\,\prime\prime\prime}_g\not\in\{x_1^{\rhd},\cdots,x_j^{\rhd}\}.$\footnote{{  This argument is subtle: if some item $y$ of $(A\setminus \{x^{\rhd^{\prime\prime\prime}}_{\vert D+1\vert},\cdots, x^{\rhd^{\prime\prime\prime}}_{\vert X\vert}\})$ would belong to $\{x_1^{\rhd},\cdots,x_j^{\rhd}\}$, then by Lemma~\ref{LEM:switches_needed_in_the_first_j_positions} there should be a reversal in which $y$ is selected, implying that $y\in D$, which is false.}}
Since $x^{\rhd^{\prime\prime\prime}}_g\not\in\{x_1^{\rhd},\cdots,x_j^{\rhd}\},$  and $c(A)=x^{\rhd^{\prime\prime\prime}}_g,$ Lemma~\ref{LEM:self_punishment_j_successors_empty} yields $x^{\rhd^{\prime\prime\prime}}_g\rhd x $ for any $x\in \{x^{\rhd^{\prime\prime\prime}}_{\vert D+1\vert},\cdots, x^{\rhd^{\prime\prime\prime}}_{\vert X\vert}\}\setminus x^{\rhd^{\prime\prime\prime}}_g.$
The definition of $\rhd^{\prime\prime\prime}$ implies that $x^{\rhd^{\prime\prime\prime}}_g\rhd^{\prime\prime\prime}x$ for any $x\in \{x^{\rhd^{\prime\prime\prime}}_{\vert D+1\vert},\cdots, x^{\rhd^{\prime\prime\prime}}_{\vert X\vert}\}\setminus x^{\rhd^{\prime\prime\prime}}_g.$
Definition~\ref{DEF:preference_compromise} implies that $c(A)=\max\left(A,\rhd^{\prime\prime\prime}_{\vert D\vert}\right).$
Since $\{\rhd^{\prime\prime\prime},\rhd^{\prime\prime\prime}_1,\cdots,\rhd^{\prime\prime\prime}_{\vert D\vert}\}$ is a rationalization by self-punishment of $c$ by $\rhd^{\prime\prime\prime}$ and $\vert D\vert< j,$ I conclude that $irr_{\mathsf{Comp}}(c)<j,$ a contradiction.

Assume now that (ii) is true. 
Thus, at least one of the conditions (ii)(a) and (ii)(b) are true also for $\{x^{\rhd}_1,\cdots x^{\rhd}_j\}.$
If condition (ii)(a) holds, then Lemma~\ref{LEM:no_external_switches} yields $irr_{\mathsf{Comp}}(c)<j,$ a contradiction.
Thus assume that (ii)(b) is true.
	There is $h\in\{1,\cdots,j\}$ satisfying one and only one of the following conditions:  
	\begin{description}

     \item[(ii)(b)(1)] there is no reversal $(A,B)$ such that either $x^{\rhd}_h=c(A)$ or $x^{\rhd}_h=c(B),$
		\item[(iii)(b)(2)] for any reversal $(A,B)$ such that either $x^{\rhd}_h=c(A)$, $y=c(B)$ or $c(A)=y,x^{\rhd}_h=c(B)$ hold, I have that  $y\in \{x^{\rhd}_1,\cdots,x^{\rhd}_{j}\}$.
\end{description} 

 If (ii)(b)(1) holds, then Lemma~\ref{LEM:switches_needed_in_the_first_j_positions} implies that $irr_{\mathsf{Comp}}(c)< j,$ a contradiction.
Similarly, when condition (ii)(b)(2) {  holds}, Lemma~\ref{LEM:switches_needed_in_the_first_j_positions} yields $irr_{\mathsf{Comp}}(c)< j$, a contradiction. 

({\bf\textit{If part}}). Assume that $c\colon\X\to X$ violates WARP under constant nonreciprocal selection of $j$ items.
Let $\rhd^c$ be the binary relation defined by
\begin{itemize}
	\item 
 $x_{g}\rhd^c x_h$ for any $g,h\in\{1,\cdots,j\}$ such that $g<h,$
 \item $y\rhd^c z$ for any $y,z\in X\setminus \{1,\cdots,j\}$ for which there is a menu $A\in\X$ such that $z\in A$ and $y=c(A),$ and
 \item $x_g\rhd^c y$ for any $g\in\{1,\cdots,j\}$ and $y\in X\setminus \{1,\cdots,j\}.$
 \end{itemize}

The binary relation $\rhd^c$ is asymmetric.
To see why, assume toward a contradiction that there are $y,z\in X$ such that $y\,\rhd^c\, z$ and $z\,\rhd^c\, y.$
Without loss of generality, one and only one of the following three cases is possibile:
\begin{enumerate}[\rm(i)]
	\item $y,z \in \{x_1,\cdots,x_j\},$
	\item $y,z\in X\setminus \{x_1,\cdots x_j\},$
	\item $y\in \{x_1,\cdots,x_j\}$ and $z\in  X\setminus \{x_1,\cdots x_j\}.$
\end{enumerate}

If (i) holds, then the definition of $\rhd^c$ implies that there are $g,h\in\{1,\cdots,j\}$ such that $g>h$ and $h>g$, a contradiction.
If (ii) holds, then there is a reversal $(A,B)$ such that either $y=c(A)$ and $z=c(B)$ or $z=c(A)$ and $y=c(B).$
Thus, the set $\{x_1,\cdots,x_j\}$ does not satisfy condition (ii)(a) of Definition~\ref{DEF:violation_independent_selection_i_items}, a contradiction.
If (iii) holds, then { the} definition of $\rhd^c$ implies that $y,z\in \{x_1,\cdots,x_j\}$ and $y,z\in X\setminus\{x_1,\cdots,x_j\},$  a contradiction.

To show that $\rhd^c$ is transitive, assume that there are $w,y,z\in X$ such that $w\rhd^c y,$ and $y\rhd^c z.$
The definition of $\rhd^c$ implies that one and only one of the following cases must be true:
{ \begin{enumerate}[\rm i)]
	\item $w,y,z\in\{x_1,\cdots,x_j\},$
	\item $w,y\in\{x_1,\cdots,x_j\},$ and $z\in X\setminus\{x_1,\cdots,x_j\},$
	\item $w\in\{x_1,\cdots,x_j\}$ and $y,z\in X\setminus\{x_1,\cdots,x_j\},$ 
	\item $w,y,z\in X\setminus \{x_1,\cdots,x_j\}.$
\end{enumerate}}

If i) holds, then the definition of $\rhd^c$ implies that there are $f,g,h\in\{1,\cdots,j\}$ such that $f>g>h,$ which implies that $f>h.$
We apply the definition of $\rhd^c$ again to conclude that $x_f\rhd^c x_h.$
If ii) or iii) are verified, since $w\in\{x_1,\cdots,x_j\}$ and $z\in X\setminus\{x_1,\cdots,x_j\},$ the definition of $\rhd^c$ yields $w\rhd^c z.$
{  Finally, if iv) is true, the definition of $\rhd^c$ implies that there are menus $A,B\in\X$ such that $w=c(A),$ and $y\in A,$ $y=c(B),$ and $z\in B.$
 Consider the set $D=\{w,y,z\}.$
If $c(D)=y,$ then $(A,D)$ is a reversal in which $w$ and $y$ are selected, which implies that the set $\{x_1,\cdots,x_j\}$ does not satisfy condition (ii)(b) of Definition~\ref{DEF:violation_independent_selection_i_items}, a contradiction.
If $c(D)=z,$ then $(B,D)$ is a reversal in which $y$ and $z$ are selected, which implies that the set $\{x_1,\cdots,x_j\}$ does not satisfy condition (ii)(b) of Definition~\ref{DEF:violation_independent_selection_i_items}, a contradiction.
Thus, I conclude that $c(D)=w.$
The definition of $\rhd^c$ implies that $w\rhd^c z.$}

Since $\rhd^c$
 is asymmetric and transitive, by \cite{Szpilrajn1930}'s theorem there is a linear order $\rhd$ that extends $\rhd^c.$
 I now show that 
 
 \begin{enumerate}[\rm(i]
 	\item
 there is  $\mathsf{Comp}_c(\rhd)\subseteq \mathsf{Comp}(\rhd)$ such that $$\max_{i\colon \rhd_i\in\mathsf{Comp}_c(\rhd)}i=j,$$ and
 \item there is no $\rhd^{\prime}\in\mathsf{LO}(X)$ such that there exists $\mathsf{Comp}_c(\rhd^{\prime})\subseteq \mathsf{Comp}(\rhd^{\prime})$ satisfying $$\max_{i\colon \rhd_i\in\mathsf{Comp}_c(\rhd^{\prime})}i<j.$$ 
\end{enumerate}

To show (i, I prove that $\{\rhd_1,\cdots,\rhd_j\}$ is a rationalization by compromise on $\rhd$ of $c$.
Consider a menu $A\in\X.$
If $c(A)\in X\setminus\{x_1,\cdots,x_j\},$ and $A\cap \setminus\{x_1,\cdots,x_j\}=\es$ then the definition of $\rhd^c$ yields $c(A)\rhd^c y$ for any $y\in A$.
Since $\rhd$ extends $\rhd^c,$ I conclude that  $c(A)=\max(A,\rhd).$
   If $c(A)\in X\setminus\{x_1,\cdots,x_j\},$ and $A\cap \setminus\{x_1,\cdots,x_j\}\neq \es$ then the definition of $\rhd^c$ implies that $c(A)\rhd^c y$ for any $y\in A\setminus\{x_1,\cdots,x_j\},$
   and $x_g\rhd^c c(A)$ for  $g\in\{1,\cdots,j\}.$
The fact that $\rhd$ extends $\rhd^c$ and Definition \ref{DEF:preference_compromise} yields $c(A)=\max(A,\rhd_j).$
If $c(A)\in \{x_1,\cdots,x_j\},$ then $c(A)=x_g$ for some $g\in\{1,\cdots,j\}.$
The definition of $\rhd_c,$ the fact that $\rhd$ extends $\rhd^c$, and Definition~\ref{DEF:preference_compromise} imply $c(A)=\max(A,\rhd_{g-1}).$ 

To show (ii, assume toward a contradiction that there is $\rhd^{\prime}\in\mathsf{LO}(\rhd),$ and a rationalization by compromise on $\rhd$ of $c$, namely $\mathsf{Comp}_c(\rhd^{\prime})$ such that $\max_{i\colon \rhd_i\in\mathsf{Comp}_c(\rhd^{\prime})}i=h<j.$
Assume $\rhd^{\prime}=\rhd.$ 
Since $c$ violates WARP under constant nonreciprocal selection of $j$ items, and the set $\{x_1,\cdots,x_j\}$ satisfies property (iii)(b) of Definition \ref{DEF:violation_independent_selection_i_items}, there is $i\in\{h+1,\cdots,j\}$ and a reversal $(A,B)$ such that either $c(A)=x_{i}$ and $c(B)=y_i$ or $c(A)=y_i$ and $c(B)=x_{i}$ for some $y_i\in X\setminus\{x_1,\cdots,x_j\}.$
 By Lemma \ref{LEM:no_external_switches} I conclude that $\mathsf{Comp}_c(\rhd^{\prime})$ is not a rationalization by compromise on $\rhd^{\prime}$ of $c$, or $h\geq j,$ a contradiction.
 Suppose now that $\rhd^{\prime}\neq \rhd.$
 Let $\{x^{\rhd^{\prime}}_1,\cdots,x^{\rhd^{\prime}}_h\}$ be the first $h$ items  of $X$ with respect to $\rhd^{\prime}.$
 Since $\mathsf{Comp}_c(\rhd^{\prime})$ is a rationalization by compromise on $\rhd^{\prime}$ of $c$ and $h=\max_{i\colon \rhd_i\in\mathsf{Comp}_c(\rhd^{\prime})}i$, Lemma~\ref{LEM:no_external_switches} implies that there is no reversal $(A,B)$ such that $c(A)\neq x^{\rhd}_{g}\neq c(B)$ for any $g\in\{0,\cdots,h\}.$
Thus, condition (ii)(a) of Definition \ref{DEF:violation_independent_selection_i_items} does not hold, and $c$ does not violate WARP under constant nonreciprocal selection of $j$ items, which is false.
\qed

\smallskip

		\smallskip
	
\noindent {\textbf{\large Proof of Theorem~\ref{THM:identification_preferences_irrational_choices_with_minimal_compromise}.}
We need some preliminary results.

\begin{lemma}\label{LEM:maximal_item_constantly_selected}
 If $c$ is an irrational choice with minimal compromise and $\mathsf{Comp}_c(\rhd)$ is a rationalization by compromise on $\rhd$ of $c$  such that $\max_{i\colon \rhd_i\in\mathsf{Comp}_c(\rhd)}i=1,$ then $\max(X,\rhd)$ is an item such that, for any reversal $(A,B),$ either $c(A)=\max(X,\rhd),$ or $c(B)=\max(X,\rhd).$ 
\end{lemma}
\begin{proof}
	Consider any pair of menus $A,B\in\X$ such that $(A,B)$ is a reversal.
	{  Two cases are possible: i) $c(A)=\max(A,\rhd)$, and $c(B)=\max\left(B,\rhd_{1}\right)$, or ii) $c(A)=\max(A,\rhd_1)$, and $c(B)=\max\left(B,\rhd\right)$.}

If case i) holds, since $(A,B)$ is a reversal, by Definition~\ref{DEF:minimal_violations_of_alpha} I have that $c(A)\neq c(B)$ and $c(A),c(B)\in (A\cap B)$.
Definition~\ref{DEF:irrational_choices_with_minimal_compromise} yields  $y\rhd_{1} \max(X,\rhd)$ for any $y\in X\setminus{\max(X,\rhd)}$, and for any $y,z\in X\setminus{\max(X,\rhd)}$ I have that  $y\rhd_{1}z$ holds only if $y\rhd z$.
	 	 Thus,  $\max(X,\rhd)\in A$ and $c(A)=\max(X,\rhd)$ hold.
	 Using the same argument of case i) in case ii), I conclude that $\max(X,\rhd)\in B$, and $c(B)=\max(X,\rhd)$.	
\end{proof}

\begin{corollary}\label{COR:irrational_choices_minimal_compromise_revealed_preference}
	Let $c\colon \X\to X$ be a choice that violates WARP under constant selection.
	y $j\in J$, either $x_j^{*}=c(A)$ or $x_j^*=c(B)$ holds.
		A pair $\left(\rhd^{c,x^*_j},\rhd^{c,x^*_j}_1\right)$ is a rationalization by self-punishment of $c$  by $\rhd^{c,x^*_j}$.

\end{corollary}

Corollary~\ref{COR:irrational_choices_minimal_compromise_revealed_preference} is an immediate consequence of Corollary~\ref{COR:multiple_minimal_preferences}.
I am now ready to prove what is left of Theorem~\ref{THM:identification_preferences_irrational_choices_with_minimal_compromise}. 
Thus, assume toward a contradiction that there is $\rhd\not \in\{\rhd^{c,x^*_j}\}_{j\in J}$, and a rationalization by compromise on $\rhd$ of $c$,	 namely $\mathsf{Comp}_c(\rhd),$ such that $\max_{i\colon \rhd_i\in\mathsf{Comp}_c(\rhd)}i=1$. 
Since $\rhd\not\in\{\rhd^{c,x^*_j}\}_{j\in J},$ for any $j\in J$  there are $w,y\in X$ such that  $w\rhd^{c,x^*_j} y$ and $y\rhd w$.
Note that the definition of $\rhd^{c,x^*_j},$ which puts $x^*_j$ on top of $X,$ implies that $y\neq x^*_j.$

Fix $j\in J.$
If $w\neq x^{*}_j ,$ and $w,y\neq x^{*}_h$ for  $h\in J\setminus\{j\}$  then, by Lemma~\ref{LEM:maximal_item_constantly_selected}, I have that $y=x^{\rhd}_i$ for some $i\in\{2,\cdots,\vert X\vert\}$, and  $w=x^{\rhd}_k$ for some $k\in\{i+1,\cdots,\vert X\vert\}.$
Thus, Lemma~\ref{LEM:necessary_condition_compromises} yields $y\rhd_1 w.$ 
Similarly, Lemma~\ref{LEM:necessary_condition_compromises} implies that $w\rhd^{c,x^*_j}_1 y.$
 Since, by Corollary \ref{COR:irrational_choices_minimal_compromise_revealed_preference},  $\left(\rhd^{c,x^*_j},\rhd^{c,x^*_j}_1\right)$ is rationalization by compromise on $\rhd^{c,x^*_j}$ of $c$,  $w\rhd^{c,x^*} y$ and $w\rhd^{c,x^*}_1 y,$ by Definition~\ref{DEF:rationalization_by_compromise} I conclude that $c(wy)=w.$
Definition~\ref{DEF:rationalization_by_compromise}, $c(wy)=w,$ $y\rhd w,$ and $y\rhd_1 w$ imply that
$\max_{i\colon \rhd_i\in\mathsf{Comp}_c(\rhd)}i>1$, a contradiction.  

{  If $w\neq x^{*}_j ,$ but $x^{*}_h=w$ or $x^{*}_h=y$ for $h\in J\setminus\{j\},$ then by Lemma~\ref{LEM:maximal_item_constantly_selected}  two cases are possible: i) $\max(X,\rhd)=x^*_j$ or ii) $\max(X,\rhd)=x^*_h.$
If i) is true, then I have that $y=x^{\rhd}_i$ for some $i\in\{2,\cdots,\vert X\vert\}$, and  $w=x^{\rhd}_k$ for some $k\in\{i+1,\cdots,\vert X\vert\}.$
Thus, Lemma~\ref{LEM:necessary_condition_compromises} yields $y\rhd_1 w.$ 
Similarly, Lemma~\ref{LEM:necessary_condition_compromises} implies that $w\rhd^{c,x^*_j}_1 y.$
 Since, by Corollary \ref{COR:irrational_choices_minimal_compromise_revealed_preference},  $\left(\rhd^{c,x^*_j},\rhd^{c,x^*_j}_1\right)$ is rationalization by compromise on $\rhd^{c,x^*_j}$ of $c$,  $w\rhd^{c,x^*} y$ and $w\rhd^{c,x^*}_1 y,$ by Definition~\ref{DEF:rationalization_by_compromise} I conclude that $c(wy)=w.$
Definition~\ref{DEF:rationalization_by_compromise}, $c(wy)=w,$ $y\rhd w,$ and $y\rhd_1 w$ imply that
$\max_{i\colon \rhd_i\in\mathsf{Comp}_c(\rhd)}i>1$, a contradiction.  
If ii) holds, then $y=\max(X,\rhd),$ and $w=x^{\rhd}_k$ for some $k\in\{2,\cdots,\vert X\vert\}.$
I now show that $\rhd\equiv \rhd^{c,x^*_h}.$
Assume toward a contradiction that there $t,v\in X$ such that $t\rhd^{c,x^*_h} v$ and $v \rhd t.$
Since $\max(X,\rhd)=\max(X,\rhd^{c,x^*_h})=x^*_h,$ I must have that $v=x^{\rhd}_i$ for some $i\in\{2,\cdots,\vert X\vert\},$ and $t=x^{\rhd}_k$ for some $k\in\{i+1,\cdots,\vert X\vert\}.$
Moreover, I must have that $t=x^{\rhd^{c,x^*_h}}_i$ for some $i\in\{2,\cdots,\vert X\vert\},$ and $v=x^{\rhd^{c,x^*_h}}_k$ for some $k\in\{i+1,\cdots,\vert X\vert\}.$
Thus, Lemma~\ref{LEM:necessary_condition_compromises} yields $v\rhd_1 t.$ 
Similarly, Lemma~\ref{LEM:necessary_condition_compromises} implies that $t\rhd^{c,x^*_j}_1 v.$
Since, by Corollary~\ref{COR:irrational_choices_minimal_compromise_revealed_preference},  $\left(\rhd^{c,x^*_h},\rhd^{c,x^*_h}_1\right)$ is {  a}  a rationalization by compromise on $\rhd^{c,x^*_h}$ of $c$,  $t\rhd^{c,x^*_h} v$ and $t\rhd^{c,x^*_h}_1 v,$ by Definition~\ref{DEF:rationalization_by_compromise} I conclude that $c(tv)=t.$
Definition~\ref{DEF:rationalization_by_compromise}, $c(tv)=t,$ $v\rhd t,$ and $v\rhd_1 t$ imply that
$\max_{i\colon \rhd_i\in\mathsf{Comp}_c(\rhd)}i>1$, a contradiction.
Thus, I obtain that $\rhd\equiv \rhd^{c,x^*_h},$ which is false.
}

If $w= x^{*}_j$, then there is some $z\in X\setminus\{x^{*}_j\}$ such that $z=\max(X,\rhd),$ and, by Lemma~\ref{LEM:maximal_item_constantly_selected}, $z=x^*_h$, with $h\in J\setminus \{j\}$.
I show that $\rhd=\rhd^{c,x^*_h}.$
To see why, toward a contradiction, if $\rhd\neq\rhd^{c,x^*_h},$ then there should be $s,t\in X$ such that $s\rhd^{c,x^*_h} t$ and $t\rhd s.$
Since $\max(X,\rhd)=\max(X,\rhd^{c,x^*_h}),$ I must have that $s,t\neq z.$
Since $s,t\neq z,$ Lemma~\ref{LEM:necessary_condition_compromises} yields $t\rhd_1 s,$ and $s\rhd^{c,z}_1 t.$
Since, by Corollary \ref{COR:irrational_choices_minimal_compromise_revealed_preference}, $(\rhd^{c,x^*_h},\rhd^{c,x^*_h}_1)$ is a rationalization by compromise on $\rhd^{c,x^*_h}$ of $c$,  $s\rhd^{c,x^*_h} t,$ and $s\rhd^{c,x^*_h}_1 t,$ Definition~\ref{DEF:rationalization_by_compromise} implies that $c(st)=s.$
Definition~\ref{DEF:rationalization_by_compromise}, $c(st)=s,$ $t\rhd s,$ and $t\rhd_1 s$ imply that
$\max_{i\colon \rhd_i\in\mathsf{Comp}_c(\rhd)}i>1$, a contradiction. 
Thus, $\rhd\equiv\rhd^{c,x^*_h},$  implying $\rhd\in \{\rhd^{c,x^*_j}\}_{j\in J},$ which is false. 
\qed

\smallskip

\noindent \textbf{\large Proof of Lemma~\ref{LEM:violation_independent_selection_i_items_equivalent_inconsistency}}.
({\bf\textit{Only if part}}). Since  $c\colon \X\to X$ violates WARP under constant nonreciprocal selection of $\vert X\vert-1,$ part (ii) of Definition~\ref{DEF:violation_independent_selection_i_items} implies that for any $D\subset X$ of cardinality $\vert X\vert-2,$ there is a reversal $(A,B)$ such that $c(A),c(B)\in X\setminus D.$
Thus, fixed arbitrary $x,y\in X,$ and considered the set $D=X\setminus\{x,y\},$ without loss of generality there is a reversal $(A,B)$ such that $c(A)=x,$ and $y=c(B).$ 
Definition~\ref{DEF:inconstistency} implies that $c$ is inconsistent.

({\bf\textit{If part}}). Straightforward.\qed

\medskip

\noindent \textbf{\large Proof of Theorem~\ref{THM:ubiquity_choices_irrationa_with_maximal_compromise}}.
By Corollary \ref{COR:equivalence_irrational_with_maximal_compromise_inconsistent_choices} it is enough to prove that the fraction of non inconsistent choices tends to 0 as the number of items in the ground set goes to infinity.
To do so, I need {  the} following notions.
\begin{definition}[\citealp{GiarlottaPetraliaWatson2022b}] \label{DEF:TFLH_properties}
	A property $\mathscr{P}$ of choices is:
	\begin{itemize}  
		\item \textsl{locally hereditary} if, when $\mathscr{P}$ holds for $c\colon \X \to X$, there are $x,y\in X$ such that, for any $Y\subseteq X$ with $x,y \in Y$, there is a choice $c^{\,\prime}\colon \mathscr{Y} \to Y$ satisfying $\mathscr{P}$;\vs
	\item \textsl{tail-fail} if, for any $k\in\mathbb{N}$, there is a set $X$ of size $\vert X \vert > k $ and a choice $c$ on $X$ such that $\mathscr{P}$ fails for any choice $c^{\,\prime}$ on $X$ satisfying $c^{\,\prime}(A)=c(A)$ for any $A \in \X$ of size at least $k$.\vs
\end{itemize}
Then $\mathscr{P}$ is a \textsl{tail-fail locally hereditary} property if it is both tail-fail and locally hereditary. Moreover, I say that $\mathscr{P}$ is \textsl{asymptotically rare} if the fraction $\frac{T(\mathscr{P},X)}{T(X)}$ of choices on $X$ satisfying $\mathscr{P}$ tends to zero as $\vert X\vert$ tends to infinity.
\end{definition}

\citet[Theorem 2]{GiarlottaPetraliaWatson2022b} prove that any tail-fail locally hereditary property of choices is asymptotically rare.
Thus, I only have to show the following:

\begin{lemma}\label{LEM:being_non_consistent_is_THFL}
	Being non inconsistent is a tail-fail locally hereditary property of choices.
\end{lemma}
\begin{proof}
To show that being non inconsistent is locally hereditary, let $c\colon\X \to X$ be a choice that is not inconsistent.
Definition~\ref{DEF:inconstistency} implies that there are distinct $x,y\in X$ for which there is no pair $A,B\in \X$ such that $c(A),c(B)\in(A\cap B)$.
The choice $c^{\,\prime}\colon \mathscr{Y}\to Y$  defined on $Y\subseteq X$ such that $x,y\in Y$, and $c(A)=c^{\,\prime}(A)$ for any $A\in \mathscr{Y}$ is non inconsistent, and this fact proves what I { am} after.

To prove that being non inconsistent is a tail-fail property, note that when $k=1$, it is enough to show that an inconsistent choice on a ground set $X$ of arbitrary size exists.
Indeed, the choice displayed in Example~\ref{EXAMPLE:existence_inconsistent_choices} is inconsistent.
Assume now that $k>1$, and let $X=\{x_{*},x_1,x_2,\cdots,x_{2k-1}\}$ be a (partially linearly ordered) ground set of cardinality $\vert X\vert=2k$.
Let $c\colon\X\to X$ be a choice such that
\begin{enumerate}[\rm(i)]

	\item  $c(X)=x_*$,
	\item $c(A)=x_j$ if $\vert A\vert=2k-1$, and $x_{j+1}\not\in A$, and
	\item $c(A)=x_j$ if $\vert A\vert=2k-2$, $x_{*}\not\in A$, and $x_{j-1}\not\in A$.
\end{enumerate}

I claim that $c$ is inconsistent.
By conditions~(i) and (ii) for any $j\leq 2k-1$ there are $B,C\in \X$ such that $x_*,x_j\in(B\cap C)$, $x_*=c(B)$, and $x_j=c(C)$.
Condition~(ii) implies that for any $i,j\leq 2k-1$ such that $j-i>1$ there are $D,E\in\X$ such that $x_i,x_j\in(D\cap E)$, $c(D)=x_i$, and $c(E)=x_j$.
Finally, conditions (ii) and (iii) imply that for any $i,j\leq 2k-1$ such that $j-i=1$ there are $F,G\in\X$ such that $x_i,x_j\in(F\cap G)$, $c(F)=x_i$, and $c(G)=x_j$.
Since $c$ is inconsistent, any choice $c^{\,\prime}$ on $X$ such that $c(A)=c^{\,\prime}(A)$ for any $A\in \X$ of cardinality $\vert A\vert\geq 2k-2\geq k$ is inconsistent.
\end{proof}

\medskip

{ \noindent \textbf{\large Proof of Theorem~\ref{THM:menu_invariant_characterization}}.
I need some preliminary lemmas.

\begin{lemma}\label{LEM:revealed_preference_follows_the_order}
	Assume that $c\colon\X\to X$ is with menu-invariant compromise of extent $i$,  $i\in\{1,\cdots,\vert X\vert-1\},$ and $(\rhd,i)$ is a  rationalization  by {  menu-invariant} compromise of $c$ for some $\rhd\in\mathsf{LO}(X).$
	If $x\,P_i\,y,$ then $x\rhd y.$
\end{lemma}

\begin{proof}
	Assume toward a contradiction that there are $x,y\in X$ such that $x\,P_i\,y$ and $y\rhd x.$
	Since $(\rhd,i)$ is a rationalization by {  menu-invariant} compromise of $c$,
 by Lemma~\ref{LEM:equivalence_menu_invariant_compromise_satisficing} $(\rhd,i)$ is a rationalization by {  menu-invariant} satisfaction of $c$.
	By Definition~\ref{DEF:P_i}  there is $A\in\X$ such that $\vert A\vert<i+1,$ $x,y\in A,$ and $y=c(A).$
	  Definition~\ref{DEF:satisficing_model} and the fact that $i\geq 1$ imply that $y=\min(A,\rhd),$ thus contradicting $y\rhd x$. 
	Thus, I conclude that $x\rhd y.$  
\end{proof}

\begin{lemma}\label{LEM:Completeness_of_P_i}
	Given a choice $c\colon\X\to X$, and $i\in\{1,\cdots,\vert X\vert-1\},$ $P_i$ is complete.
\end{lemma}

\begin{proof}
	For any $x,y\in X,$ consider the menu $xy$. If $c(xy)=y$ then Definition~\ref{DEF:P_i} implies that $x\,P_i\,y,$ if $c(xy)=x,$ then Definition~\ref{DEF:P_i} yields $y\,P_i\,x.$ I conclude that $P_i$ is complete.   
\end{proof}

Since acyclicity and completeness imply transitivity, I have 

\begin{corollary}\label{COR:P_i_linear_order_if_asymmetric_acyclic}
Consider a choice $c\colon\X\to X$, and some $i\in\{1,\cdots,\vert X\vert-1\}.$ If $P_i$ is asymmetric and acyclic, then $P_i$ is a linear order.
	
\end{corollary}

({\bf\textit{Only if part}}). Assume that $c\colon \X\to X$ is with menu-invariant {  compromise} of extent $i$, for some $i\in\{1,\cdots,\vert X\vert-1\}$, and that $(\rhd,i)$ is a rationalization by { menu-invariant} compromise of $c$ for some $\rhd\in\mathsf{LO}(X)$.
To show that $P_i$ is asymmetric, toward a contradiction assume that there are $x,y\in X$ such that $x\,P_i\,y$ and $y\,P_i\,x.$
Lemma~\ref{LEM:revealed_preference_follows_the_order} yields $x\rhd y$ and $y\rhd x,$ which is impossible, since $\rhd$ is a linear order.

To prove the acyclicity of $P_i$, assume toward a contradiction that there are (arbitrarily ordered) $x_1,\cdots,x_n\in X$ such that $x_1 P_i\cdots \cdots P_i\,x_n\,P_i\,x_1.$
Lemma~\ref{LEM:revealed_preference_follows_the_order} yields $x_1 \rhd\cdots \rhd\,x_n\,\rhd\,x_1$, which is false, since $\rhd$ is a linear order. 
Finally, to show that $P_i$ satisfies $i$-rejection, assume toward a contradiction that there is a menu $B$ of cardinality $\vert B\vert>i+1$ such that $M^{P_i^{\uparrow}}_{c(A)}\cap A\neq i.$
Since $P_i$ is asymmetric and acyclic, Corollary~\ref{COR:P_i_linear_order_if_asymmetric_acyclic} and Lemma~\ref{LEM:revealed_preference_follows_the_order} imply that $c(B)\neq \max(B\setminus B^{\rhd}_i,\rhd)$.

({\bf\textit{If part}}) 
Assume that there is some $i\in\{1,\cdots,\vert X\vert-1\}$ such that $P_i$ is asymmetric, acyclic, and it satisfies $i$-rejection.
Corollary~\ref{COR:P_i_linear_order_if_asymmetric_acyclic} implies that $P_i$ is a linear order. 
I show that $(P_i,i)$ is a rationalization by {  menu-invariant}  satisfaction of $c,$ which, by Lemma~\ref{LEM:equivalence_menu_invariant_compromise_satisficing} implies that $(P_i,i)$ is a rationalization by {  menu-invariant} compromise of $c.$ 

Consider some menu $A\in\X$ such that $\vert A\vert\leq i+.$ 
Definition~\ref{DEF:P_i} implies that $y\,P_i\,c(A)$ for any $y\in A\setminus c(A)$.
I conclude that $c(A)=\min(A,P_i).$

Consider now some menu $B\in X$ such that  $\vert B\vert> i+1.$
Since $P_i$ satisfies $i$-rejection, I have that $\left\vert M^{P_i^{\uparrow}}_{c(B)}\right\vert=i.$
Since $P_i$ is a linear order, I have that $c(B)\,P_i\,y$ for any $y\in B\setminus M^{P_i^{\uparrow}}_{c(B)}.$
I conclude that $c(B)=\max(B\setminus B^{P_i}_i,P_i).$

({\bf\textit{Uniqueness up to the extent i}}) I just proved that if $(\rhd,i)$ is a rationalization by menu-invariant compromise of $c$, $P_i$ is a linear order.
Thus, by Lemma~\ref{LEM:revealed_preference_follows_the_order} I must have that $\rhd\equiv P_i.$
\qed}

 \medskip

  \noindent \textbf{\large Proof of Theorem~\ref{THM:top_2_rationality_characterized_warp_ with the exception of two items.}.}
 ({\bf\textit{If part}})  
Assume $c\colon X\to X$ satisfies WARP with the exception of two items on top, and let $x,y\in X$ be the alternatives such that, for each reversal $(A,B)$ (if any), I have that $x,y\in A\cap B,$ and either $c(A)=x$, $c(B)=y$ or $c(A)=y$, $c(B)=x,$ and for any $A\in\X$ such that $A\cap xy,$ I have that $c(A)\in A\cap xy.$
Let $R$ be binary relation defined by $w\, R\, z$ for any $w,z\in X$ such that $\{w,z\}\neq \{x,y\},$ and there is $D\in\X$ such that $z\in B$ and $w=c(B).$
 Moreover, set either $x\,R\,y$, $\neg(y\,R\,x)$ or  $y\,R\,x$, $\neg(x\,R\,y).$  
 
 To show that $R$ is asymmetric, note that condition (ii) of Definition~\ref{DEF:WARP_exception_two_options} and the definition of $R$ imply that $x\,R\,z$, $y\,R\,z$, $\neg (z\,R\,x)$ and $\neg (z\,R\,y)$ for any $z\in X\setminus xy.$
 Moreover, by condition (i) and the definition of $R$, for any $w,z\in X\setminus xy$, I have that $w\,R\,z$ implies $\neg (z\,R\,w).$
 
 To show that $R$ is transitive, assume that there are $v,w,z\in X$ such that $v\, R\,w\,R\,z.$ 
 Since $\neg (z\,R\,x)$ and $\neg (z\,R\,xy)$ for any $z\in X\setminus xy,$ three mutually exclusive cases are possible: 1) $v=x$, $w=y$, and $z\in X\setminus xy,$ 2) $v=y,$ $w=x,$ and $z\in X\setminus xy,$ or 3) $v,w,z\in X\setminus xy.$
 If 1) or 2) holds, condition (ii) of Definition~\ref{DEF:WARP_exception_two_options} and the definition of $R$ yield $v\,R\,z.$
 If 3) holds, assume toward a contradiction $\neg(v\,R\,z).$
 Since $vz\in \X,$ I must conclude that $c(vz)=z,$ and thus $z R v.$
 However, consider the menu $vwz\in\X.$
 If $c(vwz)=v,$ then the  definition of $R$ implies that $v\,R\,z,$ contradicting asymmetry of $R.$
 If $c(vwz)=w,$ then the definition of $R$ implies that $w\,R\,z,$ contradicting asymmetry of $R.$
 Finally assume that $c(vwz)=z.$
 The definition of $R$ yields $z\,R\,w$, contradicting the asymmetry of $R.$     
 I conclude that $R$ is also transitive, and, since all the two-items menus are available, it is also complete.
 Thus $R$ is a linear order.
 
 Consider first any set $D\in\X$ such that $B\cap xy\neq xy.$
 The definition of $R$ implies $c(D)\rhd z$ for any $z\in D\setminus c(D),$ yielding $c(D)=\max(A,R).$
 Let $E\in\X$ be such that $B\cap xy= xy.$
 Condition (ii) of Definition~\ref{DEF:WARP_exception_two_options} implies that either $c(E)=x$ or $c(E)=y.$
 Thus either $c(E)=\max(A,R)$ or $c(E)=\max(A,R^{\,2 }).$
 I conclude $c$ is top-2 rationalizable. 
 
 ({\bf\textit{Only if part}}) Assume that $c$ is top-2 rationalizable, and let $\rhd\in\mathsf{LO}(X)$ be the linear order such that either $c(A)=\max(A,\rhd)$ or $c(A)=\max(A,\rhd^{2 })$ holds for any $A\in\X.$ 
 Assume toward a contradiction that there are not $a,b\in X$ such that conditions (i) and (ii) of Definition~\ref{DEF:top_2_rationality} are satisfied.
 Thus at least one of the following cases must be true.
 
 \begin{itemize}
 	\item[$(1$] There are reversals $(A,B)$ and $(D,E)$ such that $\{c(A),c(B)\}\neq \{c(D),c(E)\}$ 
 	\item[$(2$] For any $x,y\in X$ there $A\in X$ s.t. $A\,\cap xy\neq \es$ and $c(A)\not\in xy.$
 \end{itemize}
 
 If $(1$ holds, by Definition~\ref{DEF:top_2_rationality} I must have that there are at least three options $x_1,x_2,x_3\in X$ such that, for any $i,j\in\{1,2,3\}$ either $x_i\rhd x_j$, $x_j\rhd^{2 } x_i$ or $x_j\rhd x_i$, $x_i\rhd^{2 } x_j,$ which is impossible.
 If $(2$ holds, then I have that there is $A\in\X$ such that $A\,\cap\,\max(X,\rhd)\max(X\setminus \max(X,\rhd),\rhd)\neq \es$ and $c(A)\neq \max(X,\rhd),\max(X\setminus \max(X,\rhd),\rhd) $, which is impossible, since Definition~\ref{DEF:top_2_rationality} holds.
 \qed

\medskip
 \noindent \textbf{\large Proof of Lemma~\ref{LEM:uniqueness_preference_top_2_rationality}.} 
  The ``Only if part" of the proof of Theorem \ref{THM:top_2_rationality_characterized_warp_ with the exception of two items.} already showed that if $c$ satisfies WARP with the exception of two items on top, then, for any $i\in\{1,2\}$, $R_i$ is a linear order, and either $c(A)=\max(A,R_i)$ or $c(A)=\max\left(A,R_{i}^{2 }\right)$ holds for any $A\in\X.$
  Assume now that there is $\rhd^{\prime}\in\mathsf{LO}(X)$,  distinct from $R_1$ and $R_2$ such that, for any $A\in\X$, either $c(A)=\max(A,\rhd^{\prime})$ or $c(A)=\max\left(A,\rhd^{\prime\,2,\updownarrow}\right)$ holds.
  Since $\rhd^{\prime}\neq R_1$ and $\rhd^{\prime}\neq R_2$, for each $i\in\{1,2\}$ there are  $w\in X$ and $z\in X\setminus\{x,y\}$ such that $w\,R_i\,z$ and $z\,\rhd^{\prime} w$ hold.
 Since $R_i$ is a linear order, I must have that $c(wz)=w.$
 However, $w\neq \max(wz,\rhd^{\prime}),$ a contradiction.
 \qed}
 
\medskip

 {  \noindent \textbf{\large Proof of Lemma~\ref{LEM:compromise_equals_nudging}}.
Let $\rhd,\rhd^{\prime}\in\mathsf{LO}(X),$ and let $x,y\in X.$

({\bf\textit{If part}}).
Assume that $\rhd^{\prime}$ is a nudging on $-\rhd$, that is $\rhd^{\prime}\equiv -\rhd^{j}$, with $j=\vert X\vert-i$ and $i\in\{0,\cdots,\vert X\vert-1\}.$
Note that this is equivalent to assume that  $\rhd^{\prime}\equiv -\rhd^{j}$, with $j\in\{1,\cdots,\vert X\vert\}.$ 
I now prove that $\rhd^{\prime}\equiv \rhd_{i}.$
Take $a,b\in X$, and assume, without loss of generality, that $a\rhd b$.
Moreover, assume that case (i) of Definition \ref{DEF:preference_compromise} holds, that is $a,b\in X\setminus  X^{\rhd}_i$.
I obtain that $a,b\in X^{-\rhd}_j$.
Moreover, $a\rhd b$ implies $b -\rhd\,a$.
Definition~\ref{DEF:nudging} yields $a\,
\rhd^{\prime}\,b.$
Now assume that case (ii) of Definition~\ref{DEF:preference_compromise} is true, that is, $a\in X^{\rhd}_i,$ and $b\in X$.
Since $a\in X^{\rhd}_i$, I conclude that $a\not\in X^{-\rhd}_j.$
Moreover, $a\rhd b$ yields $b -\rhd\,a.$
Since $\rhd^{\prime}\equiv -\rhd^{j}$, Definition~\ref{DEF:nudging} yields $y \rhd^{\prime} x.$ 
Since  Definition~\ref{DEF:preference_compromise} holds, we conclude that $\rhd^{\prime}\equiv \rhd_i.$

({\bf\textit{Only if part}}).
Assume that $\rhd^{\prime}$ is a compromise on the preference $\rhd$,  that is $\rhd^{\prime}\equiv \rhd_{i}$, for some  $i\in\{0,\cdots,\vert X\vert-1\}.$
Let $a,b\in X$, and assume without loss of generality that $a -\rhd\,b$.
Moreover, suppose that case (i) of Definition~\ref{DEF:nudging} holds, that is, $a\in X$, and $b\in X\setminus X^{-\rhd}_j$, with $j=\vert X\vert-1$.
 The fact that $b\in X\setminus X^{-\rhd}_j$ implies that $b\in X^{\rhd}_i$.
 Moreover, $a-\rhd\,b$ implies $b \rhd a$.
Since $\rhd^{\prime}\equiv \rhd_i$, Definition~\ref{DEF:preference_compromise} yields $a \rhd^{\prime} b$.
Assume that (ii) of Definition~\ref{DEF:nudging} is true, that is, $a,b\in X^{-\rhd}_j$.
I conclude that $a,b\not\in X^{\rhd}_i$.
Moreover, $a-\rhd\,b$ implies $b\rhd a.$
Since $\rhd^{\prime}\equiv \rhd_i$, Definition~\ref{DEF:preference_compromise} implies $b\rhd^{\prime} a$.
Since Definition ~\ref{DEF:nudging} holds, we conclude that $\rhd^{\prime}=-\rhd^j$ 
\qed}

\end{document}